\documentclass[twocolumn,showpacs,preprintnumbers,nofootinbib,aps,prd,superscriptaddress,10pt]{revtex4-1}

\usepackage{setspace}
\usepackage{subcaption}  % For subfigures
\usepackage{bm,graphics,graphicx,epsfig,soul,latexsym,hyperref,mathrsfs,multirow,natbib,url}
\usepackage[usenames]{xcolor}
\usepackage{amsmath}
\usepackage{amsmath,amssymb}
\usepackage{float}
\usepackage[english]{babel}
\usepackage[normalem]{ulem}
\usepackage{caption}
\usepackage{setspace}
\usepackage{comment}
\usepackage[detect-all]{siunitx}
\usepackage{mathtools,leftindex,tensor,mhchem}
\usepackage{rotating} %for rotating the table in landscape mode
\usepackage{caption}

\definecolor{darkgreen}{rgb}{0,0.5,0}

\hypersetup{
    unicode=false,          % non-Latin characters in Acrobat’s bookmarks
    pdftoolbar=true,        % show Acrobat’s toolbar?
    pdfmenubar=true,        % show Acrobat’s menu?
    pdffitwindow=false,     % window fit to page when opened
    pdfstartview={FitH},    % fits the width of the page to the window
    pdftitle={My title},    % title
    pdfauthor={Author},     % author
    pdfsubject={Subject},   % subject of the document
    pdfcreator={Creator},   % creator of the document
    pdfproducer={Producer}, % producer of the document
    pdfkeywords={keyword1} {key2} {key3}, % list of keywords
    pdfnewwindow=true,      % links in new window
    colorlinks=true,       % false: boxed links; true: colored links
    linkcolor=red,          % color of internal links
    citecolor=blue,        % color of links to bibliography
    filecolor=magenta,      % color of file links
    urlcolor=blue,
    linktocpage=true
}

\newcommand{\TTtwo}{\texttt{TaylorT2}}
\newcommand{\TFtwo}{\texttt{TaylorF2}}
\newcommand{\TTthree}{\texttt{TaylorT3}}

\newcommand{\TTone}{\texttt{TaylorT1}}

\allowdisplaybreaks

\begin{document}

\title{5PN eccentric waveforms for intermediate-mass-ratio-inspirals (IMRIs) from post-Newtonian and black hole perturbation theory
}

% \title{Waveforms for intermediate mass ratio inspirals with nonspinning compact components on eccentric orbits through 5PN using inputs from post-Newtonian and black hole perturbation theory}

%%%%%%%%%%%%%%%%%%%%%%%%%%%%%%%%%%%%

\author{M. Laxman}
\email{mlaxman@physics.iitm.ac.in}
\affiliation{Department of Physics, Indian Institute of Technology Madras, Chennai 600036, India}
\affiliation{Centre for Strings, Gravitation and Cosmology, Department of Physics, Indian Institute of Technology Madras, Chennai 600036, India}

\author{Ryuichi Fujita}
\email{ryuichi.fujita@yukawa.kyoto-u.ac.jp}
\affiliation{Institute of General Education, Otemon Gakuin University, Osaka 567-8620, Japan}
\affiliation{Center for Gravitational Physics and Quantum Information, Yukawa Institute for Theoretical Physics, Kyoto University, Kyoto 606-8502, Japan}

\author{Chandra Kant Mishra}
\email{ckm@physics.iitm.ac.in}
\affiliation{Department of Physics, Indian Institute of Technology Madras, Chennai 600036, India}
\affiliation{Centre for Strings, Gravitation and Cosmology, Department of Physics, Indian Institute of Technology Madras, Chennai 600036, India}

%%%%%%%%%%%%%%%%%%%%%%%%%%%%%%%%%%%%
\begin{abstract}
Detection of gravitational waves from compact binaries involving at least one intermediate mass black hole, and component mass ratios in the range $0.1$--$10^{-4}$, are among the primary sources for future space detectors with target strain sensitivities in the deci-Hertz (dHz) band. Tuned to the waveform requirements for analyzing such sources, a hybrid model is obtained by combining waveforms from the post-Newtonian (PN) and black hole perturbation (BHP) theory. Components of the binary are assumed to be nonspinning and on eccentric orbits. This hybrid model is 3PN accurate in terms of results from PN theory and 5PN in results from BHP theory. In terms of eccentricity, corrections through the order $\mathcal{O}(e^{10})$ are included. Further, using number of gravitational wave cycles estimates for a few representative binaries observable in the dHz band, we demonstrate the significance of the mass ratio information from the PN approach, of contributions from the BHP theory at high PN orders, and also of higher order eccentricity corrections. In particular, we find an almost 10-fold increase in number of gravitational wave cycles for a fixed mass ratio of 0.1 and $e_0\sim0.3$ (evaluated at $0.01$Hz) when contributions beyond the leading order in eccentricity are accounted for. We also confirm the requirement to go beyond 5PN order in the circular part of the waveform from BHP theory.
\end{abstract}
%%%%%%%%%%%%%%%%%%%%%%%%%%%%%%%%%%%%
\date{\today}
\maketitle

\section{Introduction}
\label{sec:intro}
\noindent
Compact binaries with the lighter secondary (a stellar mass compact object or an intermediate mass black hole) orbiting the heavier primary, which could either be an intermediate mass black hole or a massive one, form a sub-class of compact binary source population referred to as intermediate-mass-ratio-inspirals (IMRIs)\cite{Mandel:2007hi}. These are among the primary sources for future space detectors with target sensitivities in the deci-Hertz (dHz) band~\cite{Mandel:2008bc,Amaro-Seoane:2018gbb}. Of the 200 odd events reported so far by the LIGO-Virgo-KAGRA (LVK) collaboration~\cite{LIGOScientific:2018mvr, LIGOScientific:2020ibl, LIGOScientific:2021usb, KAGRA:2021vkt, LIGOScientific:2025slb}, most have been identified as binaries involving comparable mass black holes ~\cite{KAGRA:2021duu} including events -- GW190521~\cite{LIGOScientific:2020iuh} and GW231123~\cite{LIGOScientific:2025rsn}, where the post-merger remnant is clearly an IMBH~\cite{LIGOScientific:2020ufj}. While events involving a neutron star and a black hole (BH) ~\cite{LIGOScientific:2020zkf,LIGOScientific:2025slb, KAGRA:2021vkt} could be categorized as IMRIs, none of the observed events involve pairs of a stellar mass compact object with an intermediate mass black hole (IMBH), although, such pairings seem plausible (at least by means of dynamical capture in dense stellar environments~\cite{Hong:2015aba, Haster:2016ewz, Leigh:2014oda}) and may also be detected with residual eccentricities~\cite{Antonini:2019ulv, Gondan:2020svr}, requiring accurate models accounting for the effect of eccentricity~\cite{Bhat:2022amc,Divyajyoti:2023rht, Divyajyoti:2025cwq}. With significant mass asymmetries, IMRIs spend considerable amount of time in the dHz band as opposed to their comparable component mass counterparts with similar total mass, providing a detailed map of the space-time in strong gravity regime, and thus enabling `effective' strong field tests of General Relativity~\cite{Graber:2006aa, Nakano:2021bbw, Bhat:2024hyb}. Moreover, due to this large mass asymmetry, higher order modes may be sufficiently excited to allow for consistency tests involving different modes~\cite{Dhanpal:2018ufk} apart from being useful in characterizing the source particularly when it is eccentric~\cite{Rebei:2018lzh, Chattaraj:2022tay}.

\vspace{5 pt}
While \textcolor{black}{extreme-mass-ratio-inspirals (EMRIs)} have traditionally been explored using the BHP theory techniques (see Ref.~\cite{Sasaki:2003xr, Pound:2021qin} for reviews on the subject), surrogate models calibrated to numerical relativity (NR) simulations, useful for compact binaries with mass ratios $\gtrsim10^{-4}$, are also available~\cite{Rifat:2019ltp, Islam:2022laz}, although, they still depend heavily on inputs from the BHP theory for their sensitivity \textcolor{black}{at the `small'  mass ratio end} due to 
prohibitive computational costs associated with NR simulations.\footnote{\textcolor{black}{We define the component mass ratio as, $q=m_2/m_1$, with $m_{1,2}$ representing binary constituents and $m_2<m_1$.}} Further, while these surrogates extract several higher order modes, none account for the effect of eccentricity, which, as we pointed out above, may be crucial for analyzing IMRIs. While in principle, one should not expect a good agreement between results of the BHP theory, suitable for extreme mass ratios ($<10^{-5}$), and those of NR simulations limited to mass ratios $>0.1$, identification of simple scaling relations (say for instance the total mass parameter~\cite{Rifat:2019ltp}) have allowed such comparisons and subsequent construction of these surrogates. While useful, these relations are still largely data driven and must be confirmed independently via theory or simulations. In fact, significant development has taken place in the last two decades in pushing the BHP theory under the gravitational self-force (GSF) approach to include higher order corrections in mass ratio parameter and thus improving sensitivity to mass ratios spanned by IMRIs as well as those by existing NR simulations, enabling fair comparisons between BHP theory and NR simulations; see for instance, Ref.~\cite{Albertini:2022rfe}. Additionally, the results from the BHP theory could also complement other approaches; see for instance Ref.~\cite{Albertini:2024rrs} which discusses comparisons between the GSF and effective-one-body approach. However, again, the desired second order GSF results are only available for binaries on circular orbits and thus at the moment we do not really have a theory backed model that can describe eccentric binaries with mass ratios ($q$) in the range, $0.1-10^{-4}$ --- precisely the ones that we wish to explore here.\footnote{\textcolor{black}{There have been some recent progresses for extending second order GSF calculations for non-circular orbits though; see for instance, Ref.~\cite{Mathews:2025nyb, Wei:2025lva}).}} We propose to combine the state-of-the-art results from the PN theory (see Ref.~\cite{Blanchet:2013haa} for a review) and those from the BHP theory (also referred to as first order GSF computations) to provide a fully analytical `hybrid' prescription with an expectation that such a model shall be suitable for analyzing IMRIs. The basic idea motivating this proposal is as follows. 

\vspace{5 pt}
Since the PN prescription does not involve an expansion in a mass ratio parameter, say for instance, in symmetric mass ratio ($\eta$) defined as $\eta=m_1m_2/(m_1+m_2)^2$, where $m_{1,2}$ represent masses of individual binary components, it can be used to study a binary with arbitrary mass ratio, however, pushing these computations to high PN orders, for these results to be useful for probing highly relativistic regime of compact binary dynamics, is increasingly difficult~\cite{Bernard:2015njp,Blanchet:2023sbv}. On the other hand, while waveforms based on the BHP theory provide inputs valid in the $\eta\rightarrow0$ limit, they can be pushed to very high PN orders~\cite{Fujita:2012cm}. And thus, while incomplete separately, they may provide a ``complete" model suitable for all mass ratios. While, this may sound straightforward, and it is in the circular case, it is hardly so for the eccentric case, as the two approaches employ different parametrizations. For instance, the very definition of orbital eccentricity (hence the orbital parameters depending on it) is different in the two approaches, requiring one to connect the two through desired perturbative orders prior to performing any direct comparisons; see for instance Refs.~\cite{Arun:2007sg, Arun:2009mc,Munna:2020iju}. 

\vspace{5 pt}
The computation of the PN phase has been completed to 4.5PN order for binaries on circular orbits~\cite{Blanchet:2023sbv} and up to 3PN for those on eccentric orbits~\cite{Moore:2016qxz}.\footnote{\textcolor{black}{Results are expressed as a series in a small parameter related to binary's orbital speed ($v$); a series with highest power as $v^{2n}$ is said to be $n$PN accurate.}} On the other hand, inputs required for the computation of BHP phase through 22PN for circular case in Schwarzschild spacetime~\cite{Fujita:2012cm, Varma:2013kna}, through 12PN for spherical case in Kerr spacetime~\cite{Sago:2024mgh}, and through the 5PN order for eccentric case in Kerr spacetime~\cite{Sago:2015rpa, Fujita:2020zxe} are also available.\footnote{For Schwarzschild spacetime, the fluxes have been computed through 19PN order for eccentric case \cite{Munna:2020iju,Munna:2023vds}.} Further, in the PN approach, amplitude of the dominant mode is now known through 4PN order~\cite{Blanchet:2023sbv} for the circular case and up to 3PN for the eccentric case~\cite{Boetzel:2019nfw, Ebersold:2019kdc}. Additionally, the computations of the amplitude in the test particle limit have been pushed to 5PN and to $\mathcal{O}(e^{10})$ within the BHP framework~\cite{Fujita:2020zxe}. (Note that, $\mathcal{O}(e^{10})$ refers to corrections up to the $10^{\rm th}$ power in the eccentricity parameter, $e$.) Note also, not all results available in literature from the PN approach are accurate to this order in eccentricity. For instance, Ref.~\cite{Moore:2016qxz} writes phase with only leading eccentricity corrections $[\mathcal{O}(e^{2})]$ while Ref.~\cite{Boetzel:2019nfw, Ebersold:2019kdc} provides amplitude with corrections in eccentricity up to $\mathcal{O}(e^{6})$. 
\textcolor{black}{The current work attempts to complete the description for the amplitude and phase of waveforms up to 3PN order with all the information from the PN approach and through 5PN order with $\eta\rightarrow0$ terms from the BHP theory. While our PN phase includes corrections in the eccentricity parameter through $\mathcal{O}(e^{10})$ order, PN amplitude is only $\mathcal{O}(e^{6})$ accurate. Note however, the phase as well as the amplitude from the BHP approach are $\mathcal{O}(e^{10})$ accurate and thus complement the PN results improving their sensitivity to high frequency and high eccentricity regime.} While our focus is to be able to probe IMRI space, our model by construction provides a better alternative for studying all mass ratios compared to employing a PN or BHP model alone. Our strategy is as follows. 

\vspace{5 pt}
First, the inputs required for the computation of the phase within the PN approach are pushed to $\mathcal{O}(e^{10})$. \textcolor{black}{Next, we establish relations connecting the variables of two approaches (PN and BHP) in $\eta\rightarrow0$ limit; see Sec.~\ref{subsec:BHPtoPNconv}.} Subsequently, the BHP waveform inputs are rewritten using the PN parametrization so that they could be compared with related PN expressions (Sec.~\ref{subsec:PNfluxandphase}). It is worth noting that BHP results expressed using PN parametrization are simply the $\eta\rightarrow0$ limit of corresponding PN results and thus the combined `hybrid' model is obtained by simply supplementing the PN results with those from the BHP approach (but expressed in PN parametrization) at orders where PN information is yet to be computed. The same strategy is adopted while providing amplitude and phasing pieces of our final model; see Sec.~\ref{sec:hybrid}. Although, for the amplitude model, angles describing evolution of the binary over secular time scales in the two approaches are also connected and these relations are found by comparing the radial and azimuthal frequencies of eccentric, inspiralling orbits on equatorial plane in the two approaches; see Sec.~\ref{sec:pn_bh_amp_relations} for details. 

The phasing results presented in this work include a fully analytical time-domain phasing approximant, \TTtwo{}, and its frequency domain analogue, \TFtwo{}~\cite{Damour:2000gg}. The \TTtwo{} phasing is also used in assessing the importance of mass ratio dependent terms present in the PN results (as well as of higher order PN corrections from the BHP theory) via number of GW cycle estimates. Results are summarized in Table~\ref{table:GWcycles}-\ref{table:GWcycles2}; also shown is impact of the including higher order eccentricity corrections. It should be clear from the estimates, even though eccentric contributes significantly less to number of GW cycles compared to the circular part, leading eccentricity results significantly underestimate the importance of eccentricity. We find a 10-fold increase in the GW cycles estimates when higher order terms in the eccentricity parameter are accounted for. Finally, while the eccentric part of the phasing model presented here is 5PN accurate, we also account for contributions from BHP theory in the circular part of the model up to 12PN order; see for instance Table~\ref{table:12PNorderphase} and discussion around it in Sec.~\ref{sec:GWcycles}.      

\vspace{5 pt}
The paper is organized in the following manner. Section~\ref{sec:gwphase} begins with a quick review of inputs from the two approaches (PN and BHP) required for the computation of GW phase. Relations connecting parameters employed in the two approaches are obtained in Sec.~\ref{subsec:BHPtoPNconv}. Combined PN-BHP inputs using PN parametrization are listed in Section~\ref{subsec:PNfluxandphase}. In Section~\ref{sec:modes}, we first discuss extraction of spherical harmonic mode amplitudes in the two approaches; presented separately in Sec.~\ref{subsec:PN_modes} and Sec.~\ref{subsec:BHP_modes} and subsequently obtain relations that allow us to connect the two in Sec.~\ref{sec:pn_bh_amp_relations}. In Section~\ref{sec:hybrid}, we present our hybrid model for the phase and amplitude. In Sec.~\ref{sec:GWcycles} we assess the complementarity of the two approaches through number of GW cycles estimates. Finally, in Sec.~\ref{sec:concl} we 
summarize our findings and conclude the paper. We use units in $G=c=1$.

%%%%%%%%%%%%%%%%%%%%%%%%%%%%%%%%%%%%%%%%%%%%%%%%%%%%%%%%%%%%%%%%
%%%%%%%%%%%%%%%%%%%%%%%%%%%%%%%%%%%%%%%%%%%%%%%%%%%%%%%%%%%%%%%%

\section{GW phase : inputs from PN and BHP approaches}
\label{sec:gwphase}
It is the evolution of phase whose understanding is integral in detecting a GW signal from inspiralling compact binaries, specially when large number of orbital cycles is observed in a detector and matched filtering is used to extract signals from the noisy detector data. The phase ($\phi$) that we present here is computed under the adiabatic approximation and thus ignores oscillatory contributions over orbital and periastron precession time scales and preserves only the changes over the slower, inspiral time scale.\footnote{Frequently referred as radiation reaction or secular time-scale. In the current work we use these interchangeably.} Differential equations that predict its evolution are written using energy balance laws, requiring essentially the expression of energy $E$ and the associated flux $\mathcal{F}$. In fact, under the assumption that evolution of the phase is an adiabatic process over radiation reaction time-scales, equations take the same form as in the circular case (see for instance, \cite{Moore:2016qxz}) and they read
\begin{subequations}
\label{eq:dphi_dv}
\begin{align}
\label{eq:dphi}
    \frac{d \langle\phi\rangle}{dt} &= \frac{v^{3}}{m}\,,\\
    \label{eq:dv}
    \frac{dv}{dt} &= - \frac{\langle\mathcal{F}\rangle}{dE/dv}\,,
\end{align}
\end{subequations}
where $\langle\cdot\rangle$ denotes quantities averaged over orbital time scales and $m=m_1+m_2$ is the total mass of the binary. Different ways of solving Eq.~\eqref{eq:dphi_dv} are referred to as Taylor approximants --- labeled as \TTone{}, \TTtwo{}, \TFtwo{}, \TTthree{} etc. --- giving binary's orbital phase in either time- or in frequency-domain, and can be fully analytical, semi-analytical or numerical depending upon the method used; see for instance Sec.~VI of Ref.~\cite{Moore:2016qxz} for a detailed discussion. For our current work, we limit ourselves to the only fully analytical phasing approximant referred to as \TTtwo{} and its frequency domain analogue, \TFtwo{} obtained under the stationary phase approximation (SPA) \cite{bender1999, Damour:2000gg}.\footnote{Other (semi-analytical/numerical) approximants may be easily obtained using expressions for energy and flux presented in subsequent sections following the corresponding methods outlined in Ref.~\cite{Moore:2016qxz}.}\\

We pointed out earlier in Sec.~\ref{sec:intro} that the computation of the orbit averaged PN phase ($\langle\phi\rangle$ of Eq.~\eqref{eq:dphi}) has been completed to 4.5PN order for binaries on circular orbits~\cite{Blanchet:2023sbv} and up to 3PN for those on eccentric orbits~\cite{Moore:2016qxz}. On the other hand, inputs required for the computation of BHP phase through 22PN for circular case~\cite{Fujita:2012cm} and through the 5PN order for eccentric case in Kerr spacetime~\cite{Sago:2015rpa, Fujita:2020zxe} are also available. Here we wish to develop a coherent 5PN prescription for phase by combining the results from the two approaches. However, as was noted earlier in Sec.~\ref{sec:intro}, the two (PN and BHP) representations do not involve identical parametrizations and thus one should first establish rules that connect the variables in the two approaches so that they can be directly compared and subsequently combined. Our goal in the remainder of this section is to express the inputs (expressions for energy and the flux) from the two approaches in a form that they can be combined.

%%%%%%%%%%%%%%%%%
%%%%%%%%%%%%%%%%%

\vspace{5 pt}
In the post-Keplerian representations developed within the PN approach~\cite{Memmesheimer:2004cv,damour-1985, Damour:1988mr,1993PhLA..174..196S,1995CQGra..12..983W}, \textcolor{black}{orbit averaged quantities such as the phase $\langle\phi\rangle$ of Eq.~\eqref{eq:dphi}} are expressed in terms of just two parameters which themselves change over the radiation reaction time scales. For instance, Ref.~\cite{Moore:2016qxz} writes the PN phase as a Taylor series expansion in terms of a frequency dependent small parameter $v$ and an eccentricity parameter related to the time coordinate ($e_t$); $v$ is the same parameter that appears in Eq.~\eqref{eq:dphi_dv}
($\propto\omega_\phi^{1/3}$), where $\omega_\phi$ is the orbit averaged GW half-frequency Ref.~\cite{Blanchet:2023bwj}.\footnote{\textcolor{black}{Note that the orbit averaged GW half-frequency defined in Ref.~\cite{Blanchet:2023bwj} is same as the orbit averaged azimuthal frequency ($\langle\dot{\phi}\rangle$) through the 3.5PN order but starts to differ at 4PN; see for instance, Eq.~(6.8) of Ref.~\cite{Blanchet:2023bwj}.}}
On the other hand, in the BHP approach, the results are parametrized in terms of a pair of variables, ($v_b, e_b$; labeled with a letter `$b$' to indicate the association with BHP approach). While still related to orbital frequency and eccentricity, these differ from those used in PN approach beyond the leading order; this can be immediately verified through equations relating the two sets derived below as Eqs.~\eqref{eq:vbvrelation}-\eqref{eq:eberelation}.
Our aim in the section below is to establish rules that may be used to write BHP theory results in terms of the PN parameter pair, ($v, e_t$).

\subsection{Connecting the PN and BHP inputs}
\label{subsec:BHPtoPNconv}

The connection between the two sets, $(v_b, e_b)$ and $(v, e_t)$, facilitating writing the results of the BHP theory in terms of the PN parametrization, is established in two simple steps, and in the $\eta\rightarrow0$ limit where higher order PN corrections are known from the BHP theory.\footnote{Note that, if we intended to write results in BHP parametrization, relations connecting two parametrizations in the test particle limit will no longer be sufficient, as they would be missing mass ratio information available in PN results.} First, the parameter $v_b$ is expressed in terms of the pair, ($v, e_b$), by equating the orbit-averaged azimuthal frequency in PN theory ($\omega_{\phi}\equiv v^3$) and that in the BHP theory ($\Omega_\varphi\equiv f(v_b,e_b)$). (Note the use of different symbols, $\omega_\phi$  and $\Omega_\varphi$, to represent the azimuthal frequencies of the PN and BHP approaches respectively.) Note also, Equation~(A7) of Ref.~\cite{Sago:2015rpa} gives a 3PN accurate expression for $\Omega_{\varphi}$ in terms of BHP parameters and the 5PN version is available at~\cite{BHPC}. This result is subsequently used in writing $e_b(v, e_t)$; $v_b(v, e_b)$ together with $e_b(v, e_t)$ provides necessary transformation rules for converting BHP results in terms of the PN parameter pair, $(v, e_t)$. By inverting the series, $v^3 = f(v_b, e_b)$ that follows from equality, $\omega_{\phi}=\Omega_\varphi$ mentioned above, we obtain the expression for $v_b(v, e_b)$ in $\eta\rightarrow0$ limit which is accurate through the 5PN order in $v$ and to $\mathcal{O}(e_b^{10})$ or simply to $\mathcal{O}(v^{10}, e_b^{10})$. It reads

\begin{widetext}
\begin{align}
\label{eq:vbvrelation}
    v_{b} &= v_{}\bigg[1+ \frac{e_{ b}^{2}}{2}\bigg(1-2v^{2}-6v^{4}-24v^{6}-120v^{8}-672v^{10}  \bigg)+\frac{3e_{b}^{4}}{8}\bigg(1-4v^{2}-\frac{89}{6}v^{4}-\frac{203}{3}v^{6}-\frac{1187}{3}v^{8}-2564v^{10} \bigg) \nonumber \\
    & +\frac{5e_{b}^{6}}{16}\bigg(1-6v^{2}-26v^{4}-\frac{388}{3}v^{6}-\frac{8487}{10}v^{8}-\frac{30731}{5}v^{10} \bigg)+\frac{35e_{b}^{8}}{128}\bigg(1-8v^{2}-\frac{551}{14}v^{4}-\frac{4393}{21}v^{6}-\frac{84331}{56}v^{8} \nonumber \\
    & -\frac{167333}{14}v^{10} \bigg)+\frac{63e_{b}^{10}}{256}\bigg(1-10v^{2}-\frac{3455}{63}v^{4}-\frac{19400}{63}v^{6}-\frac{134165}{56}v^{8}-\frac{15557953}{756}v^{10} \bigg) +\mathcal{O}(v^{10}, e_b^{10})\bigg]  \,,
\end{align}
\end{widetext}
where again $\mathcal{O}(v^{10}, e_b^{10})$ indicates that contributions beyond $10^{\rm th}$ powers in $v$ and $e_b$ are neglected. 

The relation, $e_b(v, e_t)$ can be obtained by solving iteratively the equation connecting the flux expressions from the two approaches; see for instance, Sec.~IX of Ref.~\cite{Arun:2007sg} for a discussion. Note that Ref.~\cite{Arun:2007sg} gives $e_t(x\equiv v^2, e\equiv e_b)$ to 3PN order and to $\mathcal{O}(e_b^2)$ and can be inverted to obtain an expression for $e_b(v, e_t)$. However, in order to consistently transform BHP theory results we require $e_b(v, e_t)$ accurate to 5PN and to $\mathcal{O}(e_t^{10})$. While, pushing these to $\mathcal{O}(e_t^{10})$ is straightforward, only a 3PN accurate version can be obtained since the energy flux for eccentric binaries from the PN approach is only 3PN accurate. \textcolor{black}{Alternatively, one can adopt the method used in Ref.~\cite{Munna:2020iju} to obtain the desired relation. In fact, Ref.~\cite{Munna:2020iju} presents a 5PN expression for $e_t(v_b, e_b)$.\footnote{\textcolor{black}{Note that, the Ref.~\cite{Munna:2020iju} uses $1/p$ as the  expansion parameter to express the result; this is the same as the square of the parameter $v_b$~\cite{Sago:2015rpa}.}} We substitute Eq.~\eqref{eq:vbvrelation} in the expression for $e_t(v_b,e_b)$ derived in Ref.~\cite{Munna:2020iju}, and invert the resulting series to obtain $e_b(v, e_t)$.} We get for $e_b(v, e_t)$, accurate to $\mathcal{O}(v^{10}, e_t^{10})$, the following expression

\begin{widetext}
\begin{align} \nonumber
\label{eq:eberelation}
    e_{b}^{2} &= e_{t}^{2} \bigg[ 1+6 v^2+32v^4+176 v^6+1008 v^8+5920 v^{10} + \frac{7}{2}e_t^2 \bigg(v^4+\frac{160 }{7}v^6+\frac{2160}{7}v^8+\frac{22688}{7}v^{10}\bigg) +\frac{13}{8} e_t^4 \bigg(v^4+\frac{514}{13} v^6\\ \nonumber
    & +\frac{40555}{52} v^8+\frac{21899}{2} v^{10}\bigg)+\frac{11}{16} e_t^6 \bigg(v^4+\frac{806}{11}v^6+\frac{21735}{11} v^8+\frac{1135292
   }{33}v^{10} \bigg)+ \frac{13}{128} e_t^{8} \bigg( v^4+\frac{4642}{13} v^6+\frac{682905}{52} v^8\\ %\nonumber
   & +\frac{10604917
   }{39}v^{10}\bigg)\bigg] + \mathcal{O}(v^{10}, e_t^{10}) \,.
\end{align}
\end{widetext}

%%%%%%%%%%%%%%%%%
%%%%%%%%%%%%%%%%%

\subsection{Conserved energy and radiated energy flux in terms of PN parameters}
\label{subsec:PNfluxandphase}
With Eqs.~\eqref{eq:vbvrelation}-\eqref{eq:eberelation}, we can now express results from the BHP theory required for the computation of the phase in terms of PN parameters and subsequently combine them with related expressions from PN approach. As indicated above, eccentric PN inputs are available through the 3PN order while those from BHP theory are 5PN accurate. Further, all results (including the relations connecting variables of the two approaches established above) assume eccentricity expansion to $\mathcal{O}(e_t^{10})$. With BHP theory results written in terms of PN parameters $(v, e_t)$ and PN results extended to the same order in eccentricity parameter (using the results shown in Appendix~\ref{App:A}) as in BHP theory, we are almost ready to express quantities appearing in the RHS of Eq.~\eqref{eq:dv}. One last step requires substituting for $e_t$, written purely in terms of the time parameter $t$ (or equivalently $v$), before we solve Eq.~\eqref{eq:dv}. The expression for $e_t$ in terms of $v$ can be derived by integrating the ratio of $de_t/dt$ and $dv/dt$. Note that, Ref.~\cite{Moore:2016qxz} explicitly lists the leading eccentricity results for $de_t/dt$ and $dv/dt$ through 3PN order and we update it through 5PN order and $\mathcal{O}(e_t^{10})$ by accounting for contributions from BHP theory but now written in terms of the PN parameters. Hybrid expression for $dv/dt$, $de_t/dt$, and hence $de_t/dv$ through 5PN and leading order in eccentricity is given in Appendix~\ref{App:PN_Evolution_eqn}. Following~\cite{Moore:2016qxz}, we express $e_t$ as a function of the PN parameter $v$, and a reference value of eccentricity ($e_0$) evaluated at a reference value of the PN parameter ($v_0$) through 5PN in terms of frequency dependent terms ($v, v_0$) and to $\mathcal{O}(e_0^{10})$ below\footnote{Note that, as far as the PN order counting is concerned both $v$ and $v_0$ are taken to be at the same footing. For instance at all three, ($v^2$, $v v_0$ and $v_0^2$), represent 1PN terms.}

\begin{widetext}
\begin{align}
\label{eq:str_et2e0}
    e_t^{\rm Hybrid}(v, v_0, e_0) &= e_0\bigg(\frac{v_0}{v}  \bigg)^{19/6}
    \frac{\mathcal{E}(v)}{\mathcal{E}(v_0)} 
    +\cdots+ \mathcal{O}(e_0^{10}) ,
\end{align}
with
\begin{align} \nonumber
\label{eq:et2e0lead}
    \mathcal{E}(v) &= 1+\bigg(-\frac{2833}{2016}+\frac{197}{72}\eta \bigg)v^2-\frac{377}{144}\pi v^3+\bigg(\frac{77006005}{24385536}-\frac{1143767}{145152}\eta +\frac{43807}{10368}\eta^2 \bigg)v^4+\bigg(\frac{9901567}{1451520}-\frac{202589}{362880}\eta \bigg)\pi v^5 \\ \nonumber
    & + \bigg[-\frac{33320661414619}{386266890240}+\frac{3317}{252} \gamma_E+\frac{180721}{41472}\pi^2+\bigg(\frac{161339510737}{8778792960}+\frac{3977}{2304}\pi^2\bigg)\eta-\frac{359037739}{20901888}\eta^2+\frac{10647791}{2239488}\eta^3 \\ \nonumber 
    & +\frac{12091}{3780} \ln 2 +\frac{26001}{1120}\ln 3 + \frac{3317}{252} \ln v \bigg]v^6-\frac{1169755952227}{122903101440}\pi v^7 + \bigg(\frac{95829848815936724147}{607396959564595200}-\frac{270890771}{5080320} \gamma_E \\ \nonumber
   & -\frac{5483124053}{418037760} \pi^2 +\frac{819986119}{15240960} \ln 2 -\frac{12157263}{100352} \ln 3-\frac{270890771}{5080320} \ln v \bigg)v^8 + \bigg(\frac{309204728020165997}{2725499177533440} -\frac{639325}{36288}\gamma_E \\ \nonumber
   & -\frac{58301801}{17915904} \pi^2-\frac{627127}{108864} \ln 2 - \frac{106893}{3584} \ln 3-\frac{639325}{36288} \ln v \bigg)\pi v^9 + \bigg(-\frac{23722481815966808777702873}{67348174876522315776000} \\  \nonumber
    & + \frac{3140631682136803}{176980466073600} \pi^2 +\frac{315895136507}{3353011200} \gamma_E -\frac{48937516919299}{90531302400} \ln 2 +\frac{17892082783}{48294400} \ln 3 + \frac{15341796875}{201180672} \ln 5 \\ 
    & + \frac{315895136507}{3353011200} \ln v \bigg)v^{10}+\mathcal{O}(v^{10}) \,,
\end{align}
\end{widetext}

\noindent
where $\gamma_E\sim0.577$ is the Euler's constant. The full expression for the evolution of $e_t$ accurate up to tenth order in eccentricity is listed in the \textbf{supplemental material}~\cite{supplementalfile}. 

Before we provide expressions for the energy and associated flux in the desired form, we wish to clarify a couple of points related to the structure of the equation and our notation here. First, we use a label, `Hybrid', to indicate that the expression has been obtained by suitably combining the information from the two theories. We argued earlier, that BHP results expressed using PN parametrization are simply the $\eta\rightarrow0$ limit of corresponding PN results and thus the combined `hybrid' model is obtained by simply supplementing the PN results with those from the BHP approach (but expressed in PN parametrization) at orders where PN information is absent. In fact, this can also be seen from Eq.~\eqref{eq:et2e0lead} which has $\eta$-independent terms beyond 3PN indicating their origin in BHP approach. 

\vspace{5pt}
With $e_t(e_0, v_0, v)$ we can explicitly write expressions for the $E$ and $\mathcal{F}$ as explicit functions of $v$ which can be substituted in the RHS of Eq.~\eqref{eq:dv} whose inverse upon performing an integral over $v$ gives $t(v)$. These are provided in Sec.~\ref{sec:hybTT2phase}. Here we merely provide expression for the flux, energy and its derivative as explicit functions of $(e_0, v_0, v)$ following Ref.~\cite{Moore:2016qxz}. Note also, we only list contributions at orders which are not explicitly listed in the literature or are not available in terms of parameters utilized here. Nevertheless, complete expressions are included as a \textbf{supplemental material}~\cite{supplementalfile}.

%%%%%%%%%%%%%%%%%%%%%%%%%%%%%

\begin{widetext}
\label{eq:Hybridflux}

\begin{align}
\label{eq:Hybridflux_circ_ecc}
    \mathcal{F}^{\rm  Hybrid}(v, v_0, e_0)
    &=\mathcal{F}^{\rm Hybrid}_{\rm circ}+\mathcal{F}^{\rm Hybrid}_{\rm ecc}\,,
\end{align}
where
\begin{align}
\label{eq:circhybridflux}
    \mathcal{F}^{\rm Hybrid}_{\rm circ}
    &=\mathcal{F}^{\rm Newt}_{\rm circ}+\cdots+\mathcal{F}^{\rm 4PN}_{\rm circ}+\mathcal{F}^{\rm 4.5PN}_{\rm circ}+\mathcal{F}^{\rm 5PN}_{\rm circ} \,,
\end{align}
with
\begin{subequations}
\label{eq:hybridfluxcirc_PNcoeff}
\begin{align} 
    \mathcal{F}_{\mathrm{circ}}^{\rm Newt} &=\frac{32}{5}v^{10}\eta^2\,,\\ \nonumber
    \mathcal{F}_{\mathrm{circ}}^{4\mathrm{PN}} &= \mathcal{F}^{\rm Newt}_{\rm circ}\bigg[-\frac{319927174267}{3178375200}-\frac{1369 }{126}\pi^2+\frac{232597 }{4410} \gamma_E+ \bigg(-\frac{1452202403629}{1466942400}-\frac{267127}{4608} \pi^2 + \frac{41478}{245} \gamma_E \\ \nonumber
    &  + \frac{479062 }{2205}\ln 2+\frac{47385}{392} \ln 3+\frac{41478}{245} \ln v\bigg)\eta+\bigg(\frac{1607125}{6804}-\frac{3157}{384} \pi ^2\bigg) \eta ^2+\frac{6875}{504} \eta^3+\frac{5}{6} \eta^4 + \frac{39931}{294} \ln 2\\
    &  -\frac{47385}{1568} \ln 3  +\frac{232597}{4410} \ln v \bigg]v^8\,, \\ \nonumber
    \mathcal{F}_{\mathrm{circ}}^{4.5\mathrm{PN}} &=  \mathcal{F}^{\rm Newt}_{\rm circ}\bigg[\frac{265978667519}{745113600}-\frac{6848}{105} \gamma_E+\bigg(\frac{2062241}{22176}+\frac{41 }{12}\pi^2\bigg)\eta -\frac{133112905}{290304} \eta ^2-\frac{3719141}{38016} \eta ^3-\frac{13696}{105} \ln 2 \\
    &  -\frac{6848}{105} \ln v\bigg]\pi v^9\,, \\ \nonumber
    \mathcal{F}_{\mathrm{circ}}^{5\mathrm{PN}} &=  \mathcal{F}^{\rm Newt}_{\rm circ}\bigg[-\frac{2489533931610883}{2831932303200}-\frac{424223}{6804} \pi ^2+\frac{916628467}{7858620} \gamma_E -\frac{83217611}{1122660} \ln 2+\frac{47385}{196}\ln 3 \\
    & +\frac{916628467}{7858620} \ln v\bigg]v^{10} \,.
\end{align}
\end{subequations}
And
\begin{align}
\label{eq:ecchybridflux}
    \mathcal{F}^{\rm Hybrid}_{\rm ecc}
    &=\mathcal{F}^{\rm Newt}_{\rm ecc}+\cdots+\mathcal{F}^{\rm 3.5PN}_{\rm ecc}+\mathcal{F}^{\rm 4PN}_{\rm ecc}
    +\mathcal{F}^{\rm 4.5PN}_{\rm ecc}+\mathcal{F}^{\rm 5PN}_{\rm ecc} \,,
\end{align}
with
\begin{subequations}
\label{eq:hybridfluxecc_PNcoeff}
\begin{align} 
    \mathcal{F}_{\mathrm ecc}^{\rm Newt} &=\mathcal{F}_{\rm circ}^{\rm Newt}\times\frac{157}{24}e_{0}^2\bigg(\frac{v_0}{v}\bigg)^{19/3}+\cdots+\mathcal{O}(e_0^{10})\,,\\ \nonumber
    \mathcal{F}_{\mathrm{ecc}}^{3.5\mathrm{PN}} &= \mathcal{F}_{\rm circ}^{\rm Newt}\times\frac{157}{24}e_{0}^2\bigg(\frac{v_0}{v}\bigg)^{19/3}\bigg[-\frac{6850154059621}{301496670720} \pi v^7 +\frac{502082620673}{28713968640} \pi v^5 v_0^2 -\frac{751181717273}{17228381184} \pi v^4 v_0^3 \\ 
    & -\frac{4239620803 }{4922394624}\pi  v^3 v_0^4-\frac{51543035947 }{2050997760} \pi  v^2 v_0^5 -\frac{200064064513 }{7681443840}\pi v_0^7\bigg]+\cdots+\mathcal{O}(e_0^{10}) \,, \\ \nonumber
    \mathcal{F}_{\mathrm{ecc}}^{4\mathrm{PN}} &=\mathcal{F}_{\rm circ}^{\rm Newt}\times\frac{157}{24}e_{0}^2\bigg(\frac{v_0}{v}\bigg)^{19/3}\bigg[ \bigg(\frac{2081538053860941646627}{5215072332511641600} +\frac{122260468151}{4101995520} \pi^2 +\frac{1619372399 }{398805120} \gamma_E \\ \nonumber
    & -\frac{95088738931}{1196415360} \ln 2 - \frac{18673443}{1125376} \ln 3  +\frac{1619372399}{398805120} \ln v \bigg)v^8 +  \bigg(\frac{21209621630052760271}{66859901698867200}-\frac{30060110267}{1640798208} \pi ^2\\ \nonumber
    & -\frac{7057192811}{99701280} \gamma_E -\frac{27447602657}{299103840} \ln 2 -\frac{482887683}{9847040} \ln 3 -\frac{7057192811}{99701280} \ln v \bigg)v^6 v_0^2 + \frac{66814383337}{2050997760} \pi^2 v^5 v_0^3  \\ \nonumber
    & +\frac{2377581791293699 }{729380745805824}v^4 v_0^4 + \frac{2717622193}{146499840} \pi ^2 v^3 v_0^5 + \bigg(-\frac{1787905184296250417}{3820565811363840} - \frac{8277347371}{234399744} \pi ^2 \\ \nonumber
    & +\frac{223522679 }{2848608} \gamma_E+\frac{814776217}{42729120} \ln 2 +\frac{194681043}{1406720} \ln 3+ \frac{223522679}{2848608} \ln v_0  \bigg)v^2 v_0^6 + \bigg(\frac{916726039513098761}{2372644373299200} \\
    & -\frac{7765571}{3732480} \pi ^2-\frac{1574437}{362880} \gamma_E -\frac{1025508937}{7620480} \ln 2 +\frac{11679093}{250880} \ln 3-\frac{1574437}{362880} \ln v_0  \bigg) v_0^8\bigg]+\cdots+\mathcal{O}(e_0^{10}) \,, \\ \nonumber
    \mathcal{F}_{\mathrm{ecc}}^{4.5\mathrm{PN}} &= \mathcal{F}_{\rm circ}^{\rm Newt}\times\frac{157}{24}e_{0}^2\bigg(\frac{v_0}{v}\bigg)^{19/3}\bigg[\bigg(-\frac{224931746076701641}{682243894886400}-\frac{622439537 }{351599616}\pi ^2 +\frac{197335927 }{7121520}\gamma_E \\ \nonumber
    &  -\frac{183119693}{3052080} \ln 2 +\frac{11582001}{100480} \ln 3+\frac{197335927}{7121520} \ln v\bigg) \pi v^9 -\frac{19406486450906293 }{303908644085760}\pi v^7 v_0^2 \\ \nonumber
    & + \bigg(\frac{2822459355640624999}{4775707264204800}-\frac{4000233523}{117199872} \pi^2-\frac{939132259 }{7121520} \gamma_E -\frac{3652575433}{21364560} \ln 2-\frac{64260027}{703360} \ln 3 \\ \nonumber
    & -\frac{939132259 }{7121520}\ln v\bigg) \pi v^6 v_0^3-\frac{211475675679731 }{86831041167360}\pi  v^5 v_0^4 -\frac{1524044093075569}{21707760291840} \pi v^4 v_0^5 + \bigg(\frac{8569803886917193}{24808868904960} \\ \nonumber
    & +\frac{3054979543}{117199872} \pi ^2-\frac{11785301}{203472} \gamma_E -\frac{42959323}{3052080} \ln 2 -\frac{10264617}{100480} \ln 3-\frac{11785301 }{203472}\ln v_0 \bigg)\pi  v^3 v_0^6  \\ \nonumber
    & + \frac{13481717115337531}{173662082334720} \pi  v^2 v_0^7 + \bigg(\frac{5549358074328361}{5323240581120}+\frac{22026761}{2239488} \pi ^2-\frac{1556101}{9072} \gamma_E -\frac{5269643}{136080} \ln 2  \\ 
    & -\frac{1366497 }{4480}\ln 3 -\frac{1556101}{9072} \ln v_0 \bigg)\pi v_0^9\bigg]+\cdots+\mathcal{O}(e_0^{10}) \,, \\ \nonumber
    \mathcal{F}_{\mathrm{ecc}}^{5\mathrm{PN}} &= \mathcal{F}_{\rm circ}^{\rm Newt}\times\frac{157}{24}e_{0}^2\bigg(\frac{v_0}{v}\bigg)^{19/3}\bigg[\bigg(-\frac{4908110178696323845432429}{289123610114445410304000}+\frac{3479359699967633 }{62022172262400}\pi ^2+\frac{597599729119139 }{9475609651200} \gamma_E  \\ \nonumber
    & +\frac{383301900806357791 }{198987802675200}\ln 2-\frac{13733857800731}{22057369600} \ln 3-\frac{6796416015625}{15792682752} \ln 5  +\frac{597599729119139}{9475609651200} \ln v \bigg)v^{10} \\ \nonumber
    & + \bigg(\frac{5896997306588047684894291}{5256792911171734732800}+\frac{346363906271783}{4134811484160} \pi ^2+\frac{4587682006367}{401995560960} \gamma_E-\frac{269386397391523}{1205986682880} \ln 2 \\ \nonumber
    &  -\frac{5877984891}{126042112} \ln 3+\frac{4587682006367 }{401995560960}\ln v \bigg) v^8 v_0^2 -\frac{2582508080477117}{21707760291840} \pi^2 v^7 v_0^3 \\ \nonumber
    &+\bigg(-\frac{8933428245563743821437}{202184342737374412800}  +\frac{12661227192449}{4961773780992} \pi ^2+\frac{2972468188817 }{301496670720}\gamma_E +\frac{11560846917779 }{904490012160}\ln 2\\ \nonumber
    & +\frac{7532993563}{1102868480} \ln 3 +\frac{2972468188817}{301496670720} \ln v \bigg) v^6 v_0^4 + \frac{135557168013761}{2584257177600}\pi^2 v^5 v_0^5 + \bigg(-\frac{52865460581479065916259}{40436868547474882560}\\ \nonumber
    & -\frac{244747755644017}{2480886890496} \pi^2+\frac{6609203597333}{30149667072} \gamma_E+\frac{24091613112859}{452245006080} \ln 2+\frac{213200116043}{551434240} \ln 3\\ \nonumber
    & +\frac{6609203597333 \ln v_0}{30149667072}\bigg) v^4 v_0^6 -\frac{710827621214689}{12404434452480} \pi^2 v^3 v_0^7 +  \bigg(-\frac{61775417624669186207507}{53640743991548313600}+\frac{523298532977}{84383907840} \pi ^2 \\ \nonumber
    & +\frac{106096586119}{8203991040}\gamma_E+\frac{69105970737619 }{172283811840}\ln 2 -\frac{87446559999}{630210560} \ln 3 +\frac{106096586119 }{8203991040}\ln v_0 \bigg)v^2 v_0^8 \\ \nonumber
    & + \bigg(-\frac{25027782258309739265857}{131539404055707648000}-\frac{622121304800869}{5530639564800} \pi^2 +\frac{27973461272447 }{140826470400}\gamma_E +\frac{776046310518967}{1267438233600} \ln 2 \\
    & +\frac{263064734543}{1545420800} \ln 3-\frac{15341796875}{100590336} \ln 5 + \frac{27973461272447}{140826470400} \ln v_0 \bigg) v_0^{10}\bigg]+\cdots+\mathcal{O}(e_0^{10}) \,,
\end{align}
\end{subequations}
\end{widetext}

%%%%%%%%%%%%%%%%%%%%%%%

\begin{widetext}
\label{eq:HybridEn}    

\begin{align}
\label{eq:HybridEn_circ_ecc}
    E^{\rm Hybrid}(v, v_0, e_0) &=  E_{\mathrm{circ}}^{\rm Hybrid}(v, v_0, e_0) + E_{\mathrm{ecc}}^{\rm Hybrid}(v, v_0, e_0) \,,
\end{align}
where
\begin{align}
\label{eq:circhybridEn}
    E^{\rm Hybrid}_{\rm circ}
    &= E^{\rm Newt}_{\rm circ}+\cdots+E^{\rm 4PN}_{\rm circ}+E^{\rm 5PN}_{\rm circ} \,,
\end{align}
with
\begin{subequations}
\label{eq:hybridEncirc_PNcoeff}
    \begin{align}
        E^{\rm Newt}_{\rm circ} &= -\frac{1}{2}\mu v^2 \,, \\ \nonumber
         E^{\rm 4PN}_{\rm circ} &= E^{\rm Newt}_{\rm circ}\bigg[-\frac{3969}{128}+ \bigg(-\frac{123671}{5760} +\frac{9037}{1536}\pi^2+\frac{896}{15}\gamma_E +\frac{1792}{15}\ln 2+\frac{896}{15}\ln v \bigg)\eta+\bigg(-\frac{498449}{3456}+\frac{3157}{576} \pi^2\bigg)\eta^2 \\
         & +\frac{301}{1728}\eta^3+\frac{77}{31104}\eta^4 \bigg]v^8 \,, \\
        E^{\rm 5PN}_{\rm circ} &= E^{\rm Newt}_{\rm circ}\bigg(-\frac{45927}{512}\bigg)v^{10}\,.
    \end{align}
\end{subequations}
\end{widetext}

And\footnote{Note that there is no Newtonian contribution to energy ($E$); see for instance Eq.~(6.2)~\cite{Moore:2016qxz}. The same is naturally true for $dE/dv$ presented below.}

\begin{widetext}
\begin{align}
\label{eq:ecchybridEn}
    E^{\rm Hybrid}_{\rm ecc}
    &=E^{\rm 1PN}_{\rm ecc}+\cdots+E^{\rm 3.5PN}_{\rm ecc}+E^{\rm 4PN}_{\rm ecc}
    +E^{\rm 4.5PN}_{\rm ecc}+E^{\rm 5PN}_{\rm ecc} \,,
\end{align}
where
\begin{subequations}
\label{eq:hybridEnecc_PNcoeff}
\begin{align}
    E^{\rm 1PN}_{\rm ecc} &=  E^{\rm Newt}_{\rm circ}\times e_{0}^2\bigg(\frac{v_0}{v}\bigg)^{19/3}(-2v^{2})+\cdots+\mathcal{O}(e_0^{10})\,,  \\
    E_{\rm ecc}^{3.5\mathrm{PN}} &= E^{\rm Newt}_{\rm circ}\times e_{0}^2\bigg(\frac{v_0}{v}\bigg)^{19/3}\bigg[\frac{744991}{12960}  \pi v^7 +\frac{1068041}{36288}\pi v^5 v_0^2 -\frac{2542111}{36288}\pi v^4 v_0^3-\frac{764881}{45360}\pi v^2 v_0^5\bigg]+\cdots+\mathcal{O}(e_0^{10})\,, \\ \nonumber
    E_{\rm ecc}^{4\mathrm{PN}} &= E^{\rm Newt}_{\rm circ}\times e_{0}^2\bigg(\frac{v_0}{v}\bigg)^{19/3}\bigg[\bigg(-\frac{8337936296887}{42247941120}-\frac{161425}{5184} \pi^2-\frac{3317 }{63}\gamma_E - \frac{12091}{945} \ln 2 -\frac{26001}{280} \ln 3  \\ \nonumber
    & -\frac{3317}{63}\ln v \bigg)v^8-\frac{194150677223}{768144384}v^6 v_0^2 + \frac{142129}{2592} \pi^2 v^5 v_0^3 +\frac{8046091493}{1536288768} v^4 v_0^4 + \bigg(-\frac{26531900578691}{84495882240}\\
    &  -\frac{122833}{5184} \pi^2 +\frac{3317}{63} \gamma_E +\frac{12091}{945} \ln 2 +\frac{26001}{280} \ln 3 +\frac{3317}{63} \ln v_0 \bigg)v^2 v_0^6\bigg]+\cdots+\mathcal{O}(e_0^{10}) \,, \\ \nonumber
    E_{\rm ecc}^{4.5\mathrm{PN}} &= E^{\rm Newt}_{\rm circ}\times e_{0}^2\bigg(\frac{v_0}{v}\bigg)^{19/3}\bigg[\frac{359242296403}{960180480} \pi  v^9 +\frac{2110559503}{13063680} \pi v^7 v_0^2-\frac{25836500287 }{54867456}\pi v^6 v_0^3 -\frac{449855627}{109734912} \pi v^5 v_0^4  \\ & -\frac{5157592583}{45722880}\pi v^4 v_0^5+\frac{200064064513}{3840721920}\pi v^2 v_0^7\bigg]+\cdots+\mathcal{O}(e_0^{10}) \,, \\ \nonumber
    E_{\rm ecc}^{5\mathrm{PN}} &= E^{\rm Newt}_{\rm circ}\times e_{0}^2\bigg(\frac{v_0}{v}\bigg)^{19/3}\bigg[\bigg(-\frac{17116170853856709871}{16608510613094400} -\frac{1925912183}{13063680} \pi ^2-\frac{270410459}{1270080} \gamma_E -\frac{1214612177}{3810240} \ln 2 \\ \nonumber
    & -\frac{33504867}{125440} \ln 3 -\frac{270410459}{1270080} \ln v \bigg) v^{10} + \bigg(-\frac{23621373529080871}{42585924648960}-\frac{457317025 }{5225472}\pi ^2 -\frac{9397061}{63504} \gamma_E   \\ \nonumber
    & -\frac{34253803 }{952560}\ln 2 - \frac{8184537}{31360} \ln 3 -\frac{9397061}{63504} \ln v \bigg) v^8 v_0^2+ \frac{280861607 }{933120}\pi^2 v^7 v_0^3 +\frac{81775675872581}{2322868617216} v^6 v_0^4  \\ \nonumber
    & +\frac{288360137}{3265920}\pi^2 v^5 v_0^5 + \bigg(-\frac{16264055054737583}{7742895390720}-\frac{828262919}{5225472} \pi ^2 +\frac{22366531}{63504} \gamma_E +\frac{81529613}{952560} \ln 2  \\ \nonumber
    & +\frac{19480527}{31360} \ln 3 + \frac{22366531}{63504} \ln v_0 \bigg) v^4 v_0^6 +  \bigg(-\frac{916726039513098761}{1186322186649600}+\frac{7765571 }{1866240}\pi^2+\frac{1574437}{181440} \gamma_E  \\
    & +\frac{1025508937 }{3810240}\ln 2 -\frac{11679093}{125440} \ln 3 + \frac{1574437}{181440} \ln v_0 \bigg)v^2 v_0^8\bigg]+\cdots+\mathcal{O}(e_0^{10})\,.
\end{align}
    
\end{subequations}
\end{widetext}

%%%%%%%%%%%%%%%%%%%%%%%%%%

\begin{widetext}
\label{eq:Hybrid_dEdv}
    
\begin{align}
\label{eq:Hybrid_dEdv_circ_ecc}
     \bigg(\frac{dE}{dv}\bigg)^{\rm Hybrid}(v, v_0, e_0) &=   \bigg(\frac{dE}{dv}\bigg)_{\mathrm{circ}}^{\rm Hybrid}(v, v_0, e_0) +  \bigg(\frac{dE}{dv}\bigg)_{\mathrm{ecc}}^{\rm Hybrid}(v, v_0, e_0) \,,
\end{align}
where
\begin{align}
\label{eq:circhybrid_dEdv}
    \bigg(\frac{dE}{dv}\bigg)^{\rm Hybrid}_{\rm circ}
    &= \bigg(\frac{dE}{dv}\bigg)^{\rm Newt}_{\rm circ}+\cdots+\bigg(\frac{dE}{dv}\bigg)^{\rm 4PN}_{\rm circ}+\bigg(\frac{dE}{dv}\bigg)^{\rm 5PN}_{\rm circ} \,,
\end{align}
with
\begin{subequations}
\label{eq:circhybrid_dEdv_PNcoeff}
    \begin{align}
        \bigg(\frac{dE}{dv}\bigg)^{\rm Newt}_{\rm circ} &= -\mu v \,, \\ \nonumber
        \bigg(\frac{dE}{dv}\bigg)^{\rm 4PN}_{\rm circ} &= \bigg(\frac{dE}{dv}\bigg)^{\rm Newt}_{\rm circ}\bigg[-\frac{19845}{128}+\bigg(-\frac{446323}{5760}+\frac{45185}{1536}\pi^2 +\frac{896}{3}\gamma_E + \frac{1792}{3} \ln 2+\frac{896}{3} \ln v \bigg)\eta + \bigg(-\frac{2492245}{3456} \\ 
        & +\frac{15785}{576} \pi^2 \bigg) \eta^2 + \frac{1505}{1728} \eta^3+\frac{385}{31104}\eta^4 \bigg]v^8 \,, \\
        \bigg(\frac{dE}{dv}\bigg)^{\rm 5PN}_{\rm circ} &= \bigg(\frac{dE}{dv}\bigg)^{\rm Newt}_{\rm circ}\bigg(-\frac{137781}{256}\bigg) v^{10} \,.
    \end{align}
\end{subequations}
And
\begin{align}
\label{eq:ecchybrid_dEdv}
    \bigg(\frac{dE}{dv}\bigg)^{\rm Hybrid}_{\rm ecc}
    &=\bigg(\frac{dE}{dv}\bigg)^{\rm 1PN}_{\rm ecc}+\cdots+\bigg(\frac{dE}{dv}\bigg)^{\rm 3.5PN}_{\rm ecc}+\bigg(\frac{dE}{dv}\bigg)^{\rm 4PN}_{\rm ecc}
    +\bigg(\frac{dE}{dv}\bigg)^{\rm 4.5PN}_{\rm ecc}+\bigg(\frac{dE}{dv}\bigg)^{\rm 5PN}_{\rm ecc} \,,
\end{align}
where
\begin{subequations}
\label{eq:Hybrid_dEdv_PNcoeff}
\begin{align}
    \bigg(\frac{dE}{dv}\bigg)^{\rm 1PN}_{\rm ecc} &= \bigg(\frac{dE}{dv}\bigg)^{\rm Newt}_{\rm circ}\times e_0^2 \bigg(\frac{v_0}{v}\bigg)^{19/3}\bigg[\frac{7}{3}v^{2}\bigg] +\cdots+\mathcal{O}(e_0^{10}) \,, \\ \nonumber
    \bigg(\frac{dE}{dv}\bigg)_{\rm ecc}^{3.5\mathrm{PN}} &=  \bigg(\frac{dE}{dv}\bigg)^{\rm Newt}_{\rm circ}\times e_0^2 \bigg(\frac{v_0}{v}\bigg)^{19/3}\bigg[\frac{744991}{9720} \pi v^7+\frac{1068041}{108864} \pi v^5 v_0^2 + \frac{2542111}{217728}\pi v^4 v_0^3+\frac{764881}{38880} \pi v^2 v_0^5\bigg] \\
    & +\cdots+\mathcal{O}(e_0^{10}) \,, \\ \nonumber
    \bigg(\frac{dE}{dv}\bigg)_{\rm ecc}^{4\mathrm{PN}} &= \bigg(\frac{dE}{dv}\bigg)^{\rm Newt}_{\rm circ}\times e_0^2 \bigg(\frac{v_0}{v}\bigg)^{19/3}\bigg[-\bigg(\frac{8944587468727}{23044331520}+\frac{1775675}{31104} \pi^2 +\frac{36487 }{378} \gamma_E+\frac{133001}{5670} \ln 2 +\frac{95337}{560} \ln 3  \\ \nonumber
    & + \frac{36487}{378} \ln v \bigg)v^8 -\frac{970753386115 }{4608866304}v^6 v_0^2 +\frac{142129}{7776} \pi^2 v^5 v_0^3 -\frac{8046091493}{9217732608} v^4 v_0^4 + \bigg(\frac{26531900578691}{72425041920}  \\
    & +\frac{859831}{31104} \pi^2 -\frac{3317}{54} \gamma_E -\frac{12091}{810} \ln 2 -\frac{8667 }{80}\ln 3 -\frac{3317}{54} \ln v_0 \bigg)v^2 v_0^6\bigg]+\cdots+\mathcal{O}(e_0^{10}) \,, \\ \nonumber
     \bigg(\frac{dE}{dv}\bigg)_{\rm ecc}^{4.5\mathrm{PN}} &=  \bigg(\frac{dE}{dv}\bigg)^{\rm Newt}_{\rm circ}\times e_0^2 \bigg(\frac{v_0}{v}\bigg)^{19/3}\bigg[\frac{359242296403 }{411505920}\pi v^9 + \frac{2110559503}{9797760}\pi v^7 v_0^2-\frac{129182501435}{329204736}\pi v^6 v_0^3  \\
    & -\frac{449855627}{329204736}\pi v^5 v_0^4 + \frac{5157592583}{274337280}\pi v^4 v_0^5 - \frac{200064064513 }{3292047360}\pi v^2 v_0^7 \bigg]+\cdots+\mathcal{O}(e_0^{10}) \,, \\ \nonumber
    \bigg(\frac{dE}{dv}\bigg)_{\rm ecc}^{5\mathrm{PN}} &=  \bigg(\frac{dE}{dv}\bigg)^{\rm Newt}_{\rm circ}\times e_0^2 \bigg(\frac{v_0}{v}\bigg)^{19/3}\bigg[\bigg(-\frac{301583169297766515167}{99651063678566400}-\frac{32740507111}{78382080} \pi^2-\frac{4596977803}{7620480}\gamma_E \\ \nonumber
    & - \frac{20648407009}{22861440} \ln 2 -\frac{189860913}{250880} \ln 3 -\frac{4596977803}{7620480} \ln v \bigg)v^{10} - \bigg(\frac{25340016298903591}{23228686172160}+\frac{5030487275}{31352832} \pi^2 \\ \nonumber
    & +\frac{103367671}{381024} \gamma_E +\frac{376791833}{5715360} \ln 2+\frac{30009969}{62720} \ln 3 +\frac{103367671}{381024} \ln v \bigg) v^8 v_0^2  \\ \nonumber
    & + \frac{280861607}{699840}\pi^2 v^7 v_0^3+\frac{408878379362905}{13937211703296} v^6 v_0^4 +\frac{288360137}{9797760}\pi^2 v^5 v_0^5 + \bigg(\frac{16264055054737583}{46457372344320} \\ \nonumber
    & +\frac{828262919}{31352832} \pi^2-\frac{22366531 }{381024}\gamma_E - \frac{81529613}{5715360} \ln 2 - \frac{6493509}{62720} \ln 3 -\frac{22366531}{381024} \ln v_0 \bigg)v^4 v_0^6  \\ \nonumber
    & + \bigg(\frac{916726039513098761}{1016847588556800}-\frac{1574437}{155520}\gamma_E-\frac{54358997}{11197440} \pi^2-\frac{1025508937 }{3265920}\ln 2 +\frac{3893031}{35840} \ln 3 \\
    & - \frac{1574437}{155520}\ln v_0 \bigg) v^2 v_0^8\bigg] +\cdots+\mathcal{O}(e_0^{10}) \,.
\end{align}

\end{subequations}
\end{widetext}

%%%%%%%%%%%%%%%%%%%%%%%%%%%%%%%%%%%%%%%%%%%%
%%%%%%%%%%%%%%%%%%%%%%%%%%%%%%%%%%%%%%%%%%%%
%%%%%%%%%%%%%%%%%%%%%%%%%%%%%%%%%%%%

\section{Spherical harmonic modes of the gravitational waveform}
\label{sec:modes}

%%%%%%%%%%%%%
%%%%%%%%%%%%%

\subsection{The PN representation}
\label{subsec:PN_modes}
The two GW polarisations ($h_+, h_\times$) written in terms of GW modes ($h^{\ell m}$) using the spin-weighted spherical harmonics take the following form~\cite{Thorne:1980ru, Blanchet:2008je,Kidder:2007rt}

\begin{align}
\label{eq:PNpolarization}
    h_{+}^{}-ih_{\times}^{} &= \sum_{\ell=2}^{\infty}\sum_{m=-\ell}^{\ell}h^{\ell m}\,Y^{\ell m}_{-2}(\Theta,\Phi) \,,
\end{align}
where $Y^{\ell m}_{-2}$ are spin-weighted spherical harmonics of weight $-2$; see Ref.~\cite{Blanchet:2008je} for explicit relations. The $h^{\ell m}$ can be expressed as 

\begin{align}
\label{eq:hlm_def}
    h^{\ell m} = \frac{2\mu v^{2}}{r}
    \,H^{\ell m} (v,e_{t},\psi,\xi) \,,
\end{align}
with $H^{\ell m}$ further decomposed as  
\begin{align}
\label{eq:Hlm_def}
    H^{\ell m} = \sqrt{\frac{16\pi}{5}}
    \,\mathrm{e}^{-im\psi}\,\hat{H}^{\ell m} (v,e_{t},\xi) \,.
\end{align}
In the above, $\mu$ represents the reduced mass and $r$ the luminosity distance to the binary. As indicated, normalised mode amplitudes ($H^{\ell m}$) are functions of PN parameters $(v, e_t)$ and of two angles ($\psi, \xi$) describing the phase of the binary. Note that, the exponential factor containing the angle $\psi$ has been separated out following the presentation of Ref.~\cite{Blanchet:2008je}, although, the two angles are related to each other; see Refs.~\cite{Boetzel:2019nfw, Ebersold:2019kdc} for details. 

%%%%%%%%%%%%%%
%%%%%%%%%%%%%%

\subsection{The BHP theory representation}
\label{subsec:BHP_modes}
The combination, $h_+-i\,h_\times$, in BHP approach can be expressed as \citep{Sasaki:2003xr, Pound:2021qin}

\begin{align}
\label{BHP_polarisation}
    h_{+}^{}-ih_{\times}^{} &= \frac{2\mu}{r}\sum_{\ell mkn}\frac{Z^{\infty}_{\ell mkn}}{w^{2}_{mkn}} \frac{\leftindex_{-2}{S}^{a\omega_{mkn}}_{\ell m}}{\sqrt{2\pi}}\,\mathrm{e}^{-i\Phi_{mkn}
     +im\varphi_0}\,, 
\end{align}
where the functions $Z^{\infty}_{\ell mkn}$ represent Teukolsky amplitudes of the Weyl scalar ($\Psi_4$) at infinity \cite{Teukolsky:1972my,Teukolsky:1974yv}, $\leftindex_{-2} {S}^{aw_{mkn}}_{\ell m}$ the spin-weighted spheroidal harmonics of weight $-2$. The phase $\Phi_{mkn}=\omega_{mkn}\cdot(t-r^*)$ (`$\cdot$' indicates simple multiplication) is a linear combination of azimuthal, polar, and radial components of the orbital phase evaluated at a given retarded time, $(t-r^*)$, where $r^*$ is the the radial coordinate in tortoise coordinates \cite{Fujita:2010xj}, and $\varphi_0$ is an arbitrary initial phase which can be set to zero~\cite{Drasco:2005kz}. Further, the frequency, $\omega_{mkn}$ is given as 
\begin{align}
\omega_{mkn}=m\Omega_{\varphi}+k\Omega_{\theta}+n\Omega_{r}\,,
\label{eq:freq_spectrum}
\end{align}
where $\Omega_\varphi, \Omega_\theta, \Omega_r$ are azimuthal, polar and radial orbital frequencies and $m, k, n$ are integers. We refer the reader to Ref.~\cite{Isoyama:2021jjd} for finer details. Note that, for nonspinning systems, the orbits can safely be assumed to be equatorial ($\theta=\pi/2$) and thus Eq.~\eqref{eq:freq_spectrum} is reduced to\footnote{In fact, this is true as long as the binary's orbit remains fixed in a plane.} 
\begin{align}
\omega_{mn}\equiv\omega_{m0n}=m\Omega_{\varphi}+n\Omega_{r}\,,
\label{eq:freq_spectrum_equatorial}
\end{align}
where we have redefined, $\omega_{mkn}$ as $\omega_{m0n}\equiv\omega_{mn}$ by setting $k=0$ in Eq.~\eqref{eq:freq_spectrum}. With similar redefinitions for other functions appearing in Eq.~\eqref{BHP_polarisation} we can write
\begin{align}
\label{BHP_polarisation_equatorial}
    h_{+}^{}-ih_{\times}^{} &= \frac{2\mu}{r}\sum_{\ell mn}\frac{Z^{\infty}_{\ell mn}}{\omega^{2}_{mn}} \frac{\leftindex_{-2} {S}^{a\omega_{mn}}_{\ell m}}{\sqrt{2\pi}}\,\mathrm{e}^{-i\Phi_{mn}} \,.
\end{align}
 Note that, the amplitudes, $Z^{\infty}_{\ell mn}$, can explicitly be written in terms of the BHP theory parameters, ($v_b, e_b$), which in turn connect with the corresponding parameters in the PN approach, ($v, e_t$) via Eqs.~\eqref{eq:vbvrelation}-\eqref{eq:eberelation} established in Sec.~\ref{subsec:BHPtoPNconv}. Further, since the spheroidal harmonic basis functions, $\leftindex_{-2} {S}^{a\omega_{mn}}_{\ell m}$, reduce to spherical harmonic basis functions, $Y^{\ell m}_{-2}$, for the non-spinning case, it allows us to cast GW modes from the BHP theory in a form similar to Eqs.~\eqref{eq:hlm_def}-\eqref{eq:Hlm_def}. We have
\begin{align}
\label{eq:hlm_bhp_def}
    h^{\ell m}_{\rm BHP} = \frac{2\mu}{r}
    \,H^{\ell m}_{\rm BHP} (v_b,e_b,\Phi_{mn}) \,,
\end{align}
with 
\begin{align}
\label{eq:Hlm_bhp_def}
    H^{\ell m}_{\rm BHP} =\frac{1}{\sqrt{2\pi}} 
    \sum_{n}\frac{Z^{\infty}_{\ell mn}(v_b,e_b)}{\omega^{2}_{mn}} \,\mathrm{e}^{-im\Phi_{\varphi}
    -in\Phi_{r}} \,,
\end{align}
where we have decomposed the phase $\Phi_{mn}$ into a radial and azimuthal phase. It should be clear now that the GW modes in the two theories can be compared (or combined) once we know how the two phase angles appearing in Eq.~\eqref{eq:Hlm_def}, $(\psi, \xi)$,  are related to the ones in Eq.~\eqref{eq:Hlm_bhp_def}, $(\Phi_\varphi, \Phi_r)$. We establish these relations in the next section.

%%%%%%%%%%%%%%%%%%%%%%%%%
%%%%%%%%%%%%%%%%%%%%%%%

\subsection{Connecting the mode amplitudes of the PN and BHP approaches}
\label{sec:pn_bh_amp_relations}
We mentioned above, that the overall phase ($\Phi_{mn}$), is evaluated at a given retarded time which in the source frame can simply be expressed as, $\Phi_{ mn}=\omega_{ mn}\cdot\,t$.\footnote{In harmonic coordinates, $r^*$ is given as $r^*=r_{\rm H}+2M\ln((r_{\rm H}-M)/r_0)$ by setting $r_0=2M/\sqrt{\mathrm{e}}$\,, where $\rm H$ stands for harmonic gauge. Then, we may find the phase $\Phi_{mn}$ in Eq.~(\ref{BHP_polarisation_equatorial}) acquires logarithmic corrections from retarded to source frame time transformation in radial and secular azimuthal phases Eqs.~\eqref{eq:xitoell_gen} and \eqref{eq:lambdaxitolambda_gen} respectively.}
\begin{subequations}
\label{eq:BHPphaseangles}
\begin{align}
    \Phi_\varphi &= \Omega_\varphi\cdot t\equiv\Omega_\varphi t=\omega_{\phi}\,t\,,\\
    \Phi_r&=\Omega_r\cdot t\equiv\Omega_r t=\omega_r t, 
\end{align}
\end{subequations}
where we have dropped the `$\cdot$' for convenience. The two PN frequencies, ($\omega_{\phi},\omega_r$), within the post-Keplerian representation, are expressed as a function of the mean motion ($n$), and the constant $k$ characterizing the periastron precession~\cite{Moore:2016qxz}\footnote{Note that, variables ($n, k$) that appear in Eq.~\eqref{eq:PNfreq} are different from indices ($n, k$) in say Eq.~\eqref{BHP_polarisation}.} \\
\begin{subequations}
\label{eq:PNfreq}
\begin{align}
    \omega_{\phi} &= (1+k)\omega_r \,,\\
    \omega_r&=n\,.
\end{align}
\end{subequations}
Note again, $\omega_{\phi}$ is orbit-averaged azimuthal frequency, i.e $\omega_{\phi}=\langle{\dot  \phi}\rangle$.\footnote{See Eq.~(3.2f) of Ref.~\cite{Moore:2016qxz} for the expression for the non-averaged definition for azimuthal frequency (${\dot \phi}$).} Using above in Eq.~\eqref{eq:BHPphaseangles} we can write,
\begin{subequations}
\label{eq:BHPphaseangles-l}
\begin{align}
    \Phi_\varphi&=(1+k)\,n\,t=(1+k)\,l=\langle\phi\rangle\,,\\
    \Phi_r&=n\,t=l, 
\end{align}
\end{subequations}

\noindent
where $l$ is mean anomaly. It is important to note that the formal phase ($\phi$), apart from the secularly growing phase, $\langle{\phi}\rangle\equiv\lambda$, also has oscillatory contributions ($W(l)$) with amplitudes varying over radiation reaction time scales and can be expressed as\footnote{Note that, $\phi=\langle\phi\rangle(\equiv\lambda)+\tilde{\lambda}+W(l)$; see for instance, Eq.~(5.1) of Ref.~\cite{Moore:2016qxz} and the discussion around it. However, we have ignored $\tilde{\lambda}$ on account of it being $\propto\eta$ and thus vanishes in the test mass limit.} 

\begin{align}
 \label{eq:formalphi_approxm}
\phi\simeq\langle\phi\rangle+W(l)\,,
\end{align}

Using Eqs.~\eqref{eq:BHPphaseangles-l}-\eqref{eq:formalphi_approxm} in Eq.~\eqref{eq:Hlm_bhp_def} and subsequently expressing the result we have
\begin{align}
\label{eq:Hlm_bhp_def_2}
    H^{\ell m}_{\rm BHP} &= \frac{1}{\sqrt{2\pi}} 
    \sum_{n}\frac{Z^{\infty}_{\ell mn}(v_b,e_b)}{\omega^{2}_{mn}} \,\mathrm{e}^{imW(l)}\mathrm{e}^{-im\phi-inl} \,.
\end{align}
The function $W(l)$ up to 3PN order was computed in Ref.~\cite{Konigsdorffer:2006zt} and we extend it through 5PN (in the test mass limit) using the inputs from Refs.~\cite{Munna:2020iju,BHPC}. In terms of the PN variables ($e_t, v$) and mean-anomaly $l$ it reads, 
\begin{align} \nonumber
\label{eq:W(l)_function}
    W(l) &=  2e_t(1 + 5 v^2 + 26 v^4 + 146 v^6 + 854 v^8 \\ 
    & + 5078 v^{10}+\mathcal{O}(v^{10}))\sin (l)+\mathcal{O}(e_t)\,,
\end{align}
\noindent
Further, the angles, $(\phi, l)$ are related to the two phase angles, $(\psi,\xi)$ appearing in  Eq.~\eqref{eq:Hlm_def} as follows 

\begin{widetext}
\begin{subequations}
\label{eq:phi2psi_l2xi}
\begin{align}
\label{eq:phitopsi_5PNe1}
\phi&=
\psi + 6 \bigg\{v^3 + 2e_t\bigg[(v^3 + 2v^5)\cos (\xi) -3v^6 \sin (\xi) \ln \bigg(\frac{v}{v^{\prime}_0}\bigg) +\frac{13}{2}v^7\cos (\xi) + 3v^8 \sin (\xi) \ln \bigg(\frac{v}{v^{\prime}_0}\bigg)
\nonumber \\
&  + v^9\bigg(32-6\ln^2 \bigg(\frac{v}{v^{\prime}_0}\bigg)\bigg)\cos (\xi)+12v^{10}\sin (\xi) \ln \bigg(\frac{v}{v^{\prime}_0}\bigg) \bigg] +\mathcal{O}(v^{10})+
\mathcal{O}(e_t^{}) \bigg\}\ln \bigg(\frac{v}{v^{\prime}_0}\bigg)\,,
\\
\label{eq:elltoxi_5PNe1}
l&=\xi+6\bigg\{v^3-3v^5 -\frac{9}{2}v^7 -\frac{27}{2}v^9+\mathcal{O}(v^{10})+
\mathcal{O}(e_t^{}) \bigg\}\ln \bigg(\frac{v}{v^{\prime}_0}\bigg)
\,, 
\end{align}
\end{subequations}
\end{widetext}
where
\begin{align}
\label{eq:logv0p_logv0}
\ln v^\prime_0 = \frac{11}{36}-\frac{1}{3}\gamma_E - \frac{2}{3}\ln 2 +\frac{2}{3}\ln v_0.
\end{align}

Note that $v_0=\sqrt{M/r_0}$ is a freely specifiable constant and can be fixed using $r_0 = 2M/\sqrt{\mathrm{e}}$ in order to match the BHP results in Schwarzschild coordinates with PN results in harmonic coordinates under test mass limit \cite{Fujita:2010xj}. The simplified expression is given by
\begin{align}
\label{eq:logv0p_simpl}
\ln v^\prime_0 = \frac{17}{36}-\ln 2-\frac{1}{3}\gamma_E.
\end{align}
These redefinitions help remove arbitrary dependence of resulting waveform on arbitrary frequency scales ($v_0'$); see for instance~\cite{Boetzel:2017zza} for details. It should be clear that to reexpress BHP modes (Eq.~\eqref{eq:Hlm_bhp_def_2}) in terms of PN variables ($v, e_t, \xi, \psi$) we only need these results in the test mass limit. Details of these computations are outlined in Appendix \ref{App:phitopsi_ltoxi_gen_eqn}. Finally, the Teukolsky amplitudes (with $k=0$) can be computed via $Z^{\infty}_{\ell mn} = z^{\infty}_{\ell mn}\mathrm{e}^{-i \phi_{\rm binc}}$, where the expression for $z^{\infty}_{\ell mkn}$ and $\phi_{\rm binc}$ can be accessed at~\cite{BHPC}. Note that, $\omega_{mn}$ is defined in Eq.~\eqref{eq:freq_spectrum_equatorial} and explicit expressions for $\Omega_{\varphi}$ and $\Omega_{r}$ with desired accuracies can be accessed at~\cite{BHPC}. Note also, the summation over $n$ depends on the highest eccentricity order used. Since our results are $\mathcal{O}(e^{10})$ accurate we perform the sum up to $n=10$. Together with Eqs.~\eqref{eq:vbvrelation}-\eqref{eq:eberelation} these allow us to reexpress Eq.~\eqref{eq:Hlm_bhp_def_2} completely in terms of PN variables used in Eq~\eqref{eq:Hlm_def} and subsequently combine the two sets of results.

%%%%%%%%%%%%%%%%%%%%%%%%%%%%%%%%%%%%%%%%%%%%%%%%%%%%%%%%%
%%%%%%%%%%%%%%%%%%%%%%%%%%%%%%%%%%%%%%%%%%%%%%%%%%%

\section{Waveform model}
\label{sec:hybrid}
Having established relations required to connect the BHP and PN results, the hybrid PN-BHP model can now be constructed. The two approximants for the orbital phase are presented first and have the same structure as adopted in Sec.~\ref{subsec:PNfluxandphase} for expressing the inputs required for the computations of the phase.

%%%%%%%%%%%%%%%%%
%%%%%%%%%%%%%%%%%

\subsection{\TTtwo{} hybrid phase}
\label{sec:hybTT2phase}
We highlighted in Sec.~\ref{sec:gwphase} that different ways of solving Eq.~\eqref{eq:dphi_dv} are referred to as Taylor approximants and in this work we explicitly list the only time-domain approximant, \TTtwo{}, that takes a fully analytical form. With minor rearrangements to Eq.~\eqref{eq:dphi_dv} we can cast it in a form so as to express it as a set of equations involving only the PN parameter $v$ and it reads\footnote{In fact, it's not difficult to notice from this form that effectively we need to solve just one equation since the RHS of the two equation is related by a factor purely dependent on $v$.} 

\begin{subequations}
\label{eq:dphi_dv_TT2}
\begin{align}
\label{eq:dphi_TT2}
    \frac{d \langle\phi\rangle}{dv} &= \frac{d \langle\phi\rangle}{dt}\,\frac{dt}{dv}=\frac{v^{3}}{m}\,\,\frac{dt}{dv}\,,\\
    \label{eq:dv_TT2}
    \frac{dt}{dv} &= - \frac{dE(v)/dv}{\langle\mathcal{F}(v)\rangle}\,.
\end{align}
\end{subequations}

With $\mathcal{F}(v)$, and $dE/dv$ written explicitly in Sec.~\ref{subsec:PNfluxandphase}, their ratio appearing in Eq.~\eqref{eq:dv_TT2} and thus in the RHS of Eq.~\eqref{eq:dphi_TT2}, it can be fully expanded in the three small parameters, ($v$, $e_0$, $v_0$), and subsequently integrated to obtain closed-form expressions for the phase, $\langle\phi\rangle$ and $t$, as functions of the three parameters. To $\mathcal{O}(v^{10},e_0^{10})$, for $\langle\phi\rangle$ we get

\begin{widetext}
\label{eqn:hybridphase}
\begin{align}
\label{eq:phaseHybrid_circ_ecc}
    \langle\phi\rangle^{\rm Hybrid}(v, v_0, e_0) - \phi_{c} &= \langle\phi\rangle_{\mathrm{circ}}^{\rm Hybrid}(v, v_0, e_0) + \langle\phi\rangle_{\mathrm{ecc}}^{\rm Hybrid}(v, v_0, e_0) \,,
\end{align}
where
\begin{align}
\label{eq:phaseHybrid_circ}
\langle\phi\rangle_{\mathrm{circ}}^{\rm Hybrid}(v, v_0, e_0) &= \langle\phi\rangle^{\rm Newt}_{\rm circ}+\cdots+\langle\phi\rangle^{\rm 4PN}_{\rm circ}+\langle\phi\rangle^{\rm 4.5PN}_{\rm circ}+\langle\phi\rangle^{\rm 5PN}_{\rm circ} \,,
\end{align}
with
\begin{subequations}
\label{eq:phaseHybrid_circ_PNCoeff}
\begin{align} 
\langle\phi\rangle^{\rm Newt}_{\rm circ} &= -\frac{1}{32v^5\eta} \,, \\
\langle\phi\rangle_{\mathrm{circ}}^{4\mathrm{PN}} &= \langle\phi\rangle^{\rm Newt}_{\rm circ}\bigg[\frac{2554404624135128353}{2214468081745920}-\frac{9203}{126}\gamma_E -\frac{45245}{756} \pi^2+ 
\bigg(-\frac{680712846248317}{337983528960}-\frac{488986}{1323}\gamma_E \\ \nonumber
& +\frac{109295}{1792} \pi^2-\frac{1245514}{1323} \ln 2 + \frac{78975}{392} \ln 3-\frac{488986}{1323} \ln v \bigg)\eta +\bigg(\frac{7510073635}{24385536}-\frac{11275 \pi ^2}{1152}\bigg)\eta ^2 \\
& +\frac{1292395}{96768} \eta^3-\frac{5975}{768} \eta^4-\frac{252755}{2646} \ln 2-\frac{78975}{1568} \ln 3 -\frac{9203}{126} \ln v \bigg]v^8\,,\\ \nonumber
\langle\phi\rangle_{\mathrm{circ}}^{4.5\mathrm{PN}} &= \langle\phi\rangle^{\rm Newt}_{\rm circ}\bigg[-\frac{93098188434443}{150214901760}+\frac{1712}{21}\gamma_E + \frac{80}{3} \pi^2+\bigg(\frac{1492917260735}{1072963584}-\frac{2255}{48} \pi ^2 \bigg)\eta -\frac{45293335}{1016064} \eta
^2  \\
& - \frac{10323755}{1596672} \eta^3 + \frac{3424}{21} \ln 2+\frac{1712}{21} \ln v \bigg]\pi v^9\,, \\ \nonumber
\langle\phi\rangle_{\mathrm{circ}}^{5\mathrm{PN}} &= \langle\phi\rangle^{\rm Newt}_{\rm circ}\bigg[\frac{474387630222958367413}{168408930661171200}-\frac{6470582647}{110020680} \gamma_E-\frac{578223115}{12192768} \pi^2 -\frac{53992839431}{220041360} \ln 2+\frac{5512455}{87808}\ln 3 \\
& -\frac{6470582647}{110020680} \ln v \bigg] v^{10} \,. \\ \nonumber
\end{align}
\end{subequations}
And
\begin{align}
\label{eq:hybridphase_ecc}
    \langle\phi\rangle^{\rm Hybrid}_{\rm ecc}
    &= \langle\phi\rangle^{\rm Newt}_{\rm ecc}+\cdots+\langle\phi\rangle^{\rm 3.5PN}_{\rm ecc}+\langle\phi\rangle^{\rm 4PN}_{\rm ecc}
    +\langle\phi\rangle^{\rm 4.5PN}_{\rm ecc}+\langle\phi\rangle^{\rm 5PN}_{\rm ecc} \,,
\end{align}
with
\begin{subequations}
\label{eq:hybridphase_ecc_PNCoeff}
    \begin{align}
        \langle\phi\rangle^{\rm Newt}_{\rm ecc} &=  \langle\phi\rangle^{\rm Newt}_{\rm circ}\times\bigg(-\frac{785}{272}\bigg)e_0^2 \bigg(\frac{v_0}{v}\bigg)^{19/3} +\cdots+\mathcal{O}(e_0^{10}) \,, \\ \nonumber
        \langle\phi\rangle^{\rm 3.5PN}_{\rm ecc} &= \langle\phi\rangle^{\rm Newt}_{\rm circ}\times\bigg(-\frac{785}{272}\bigg)e_0^2 \bigg(\frac{v_0}{v}\bigg)^{19/3}\bigg[\frac{660212554196821}{7838913438720}\pi v^7 -\frac{186549273611}{4262229720}\pi v^5v_0^2 + \frac{142362943957}{7735191552}\pi v^4v_0^3  \\ 
        & +\frac{1329922389787}{430709529600}\pi v^3 v_0^4+\frac{5319946988941}{200997780480}\pi v^2 v_0^5-\frac{200064064513}{7681443840}\pi v_0^7\bigg] +\cdots+\mathcal{O}(e_0^{10}) \,,\\ \nonumber
        \langle\phi\rangle^{\rm 4PN}_{\rm ecc} &= \langle\phi\rangle^{\rm Newt}_{\rm circ}\times\bigg(-\frac{785}{272}\bigg)e_0^2 \bigg(\frac{v_0}{v}\bigg)^{19/3}\bigg[\bigg(-\frac{32805947446710779617}{71635608963072000} - \frac{213222313}{2540160}\gamma_E -\frac{25555945861}{640936800}\pi^2 \\ \nonumber
& - \frac{2412688001387}{5982076800}\ln 2 - \frac{22410327537}{196940800}\ln 3 - \frac{213222313}{2540160}\ln v \bigg)v^8 + \bigg(-\frac{580239664795957088617}{1069758427181875200}\\ \nonumber
& +\frac{35366596901}{797610240}\gamma_E + \frac{1035857046293 }{13126385664}\pi^2 + \frac{465424002959}{2392830720}\ln 2 -\frac{8209090611}{78776320}\ln 3 + \frac{35366596901}{797610240}\ln v \bigg) v^6 v_0^2  \\ \nonumber
& -\frac{24824947459}{304444980}\pi^2 v^5 v_0^3 - \frac{450596088168791}{327477069545472} v^4 v_0^4  - \frac{852488175097}{12818736000}\pi^2 v^3 v_0^5+ \bigg(\frac{184536293350846943351}{374415449513656320} \\ \nonumber
& -\frac{23070600737}{279163584}\gamma_E+\frac{854335574413}{22971174912}\pi^2 -\frac{84096060751}{4187453760}\ln 2 - \frac{20093749029}{137858560}\ln 3 \\ \nonumber
& -\frac{23070600737}{279163584}\ln v_0 \bigg) v^2 v_0^6 + \bigg(\frac{916726039513098761}{2372644373299200} -\frac{1574437}{362880}\gamma_E - \frac{7765571}{3732480} \pi^2 - \frac{1025508937}{7620480}\ln 2 \\ 
& + \frac{11679093}{250880}\ln 3 - \frac{1574437}{362880}\ln v_0 \bigg)v_0^8\bigg] +\cdots+\mathcal{O}(e_0^{10}) \,, \\ \nonumber
\langle\phi\rangle^{\rm 4.5PN}_{\rm ecc} &= \langle\phi\rangle^{\rm Newt}_{\rm circ}\times\bigg(-\frac{785}{272}\bigg)e_0^2 \bigg(\frac{v_0}{v}\bigg)^{19/3}\bigg[\bigg(\frac{10763550793829749591}{11700482797301760} + \frac{138576877}{712152} \gamma_E  + \frac{27007910591 }{1230598656}\pi^2 \\ \nonumber
& -\frac{64303456267}{74775960} \ln 2 + \frac{3047314311}{2461760} \ln 3 +\frac{138576877}{712152} \ln v \bigg)\pi v^9 + \frac{1870382166039593893}{7901624746229760} \pi v^7 v_0^2   \\ \nonumber
& + \bigg(-\frac{77215091291237494673}{76411316227276800} + \frac{4706391469 }{56972160} \gamma_E + \frac{137846137117 }{937598976}\pi^2 + \frac{61936056871}{170916480} \ln 2 - \frac{1092420459}{5626880} \ln 3  \\ \nonumber
& + \frac{4706391469}{56972160} \ln v \bigg)\pi v^6 v_0^3 + \frac{78573987746417}{12888982673280} \pi v^5 v_0^4 + \frac{288834777020621}{9746341355520} \pi v^4 v_0^5  \\ \nonumber
& + \bigg(-\frac{29570784875272531067 }{23878536321024000} +\frac{3696919229}{17803800} \gamma_E - \frac{136901923321}{1464998400} \pi^2 + \frac{13475866867}{267057000} \ln 2+\frac{3219897393}{8792000} \ln 3  \\ \nonumber
& + \frac{3696919229}{17803800} \ln v_0 \bigg)\pi v^3 v_0^6 -\frac{1391497785408752893}{17018884068802560} \pi v^2 v_0^7 + \bigg(\frac{5549358074328361}{5323240581120} -\frac{1556101}{9072} \gamma_E \\
& + \frac{22026761}{2239488} \pi^2 - \frac{5269643}{136080} \ln 2 - \frac{1366497}{4480} \ln 3 - \frac{1556101}{9072} \ln v_0 \bigg)\pi v_0^9 \bigg] +\cdots+\mathcal{O}(e_0^{10}) \,, \\ \nonumber
\langle\phi\rangle^{\rm 5PN}_{\rm ecc} &= \langle\phi\rangle^{\rm Newt}_{\rm circ}\times\bigg(-\frac{785}{272}\bigg)e_0^2 \bigg(\frac{v_0}{v}\bigg)^{19/3} \bigg[\bigg(\frac{271279403411232137151271099}{20651686436746100736000} - \frac{8976684795508121}{2707317043200} \gamma_E  \\ \nonumber
& -\frac{4703506315734395131}{1736620823347200} \pi^2 + \frac{536757830527233971}{56853657907200} \ln 2- \frac{421047256217977}{44114739200} \ln 3 - \frac{115539072265625}{31585365504} \ln 5  \\ \nonumber
& -\frac{8976684795508121 }{2707317043200} \ln v \bigg) v^{10} + \bigg(-\frac{92939249116531638654961}{72208693834776576000}-\frac{604058812729 }{2560481280}\gamma_E \\ \nonumber
& - \frac{72399994624213 }{646064294400}\pi^2 -\frac{6835145107929371}{6029933414400}  \ln 2 -\frac{7054273101369}{22057369600} \ln 3- \frac{604058812729 }{2560481280} \ln v \bigg)v^8 v_0^2\\ \nonumber
& + \frac{19146164071707809}{43415520583680} \pi^2 v^7 v_0^3 + \bigg( \frac{244395185406791596169899}{3234949483797990604800} - \frac{14896303254047}{2411973365760} \gamma_E \\ \nonumber
& -\frac{436299843397871}{39694190247936} \pi^2 - \frac{196035177181373}{7235920097280} \ln 2 +\frac{128060890571}{8822947840} \ln 3  - \frac{14896303254047}{2411973365760} \ln v  \bigg) v^6 v_0^4 \\ \nonumber
& -\frac{50366394263627}{383600674800} \pi^2 v^5 v_0^5+ \bigg(\frac{10018990650283508491831}{18155328735600967680} - \frac{1252567334497}{13536585216} \gamma_E + \frac{46384263912653}{1113867583488} \pi^2 \\ \nonumber
& -\frac{4565809961231 }{203048778240}\ln 2 - \frac{40405397887 }{247582720}\ln 3 - \frac{1252567334497}{13536585216} \ln v_0\bigg) v^4 v_0^6+\frac{222978802270125481 }{1085388014592000}\pi^2 v^3 v_0^7  \\ \nonumber
& +\bigg( \frac{6376068870309914801531621}{5256792911171734732800} - \frac{10950620263057}{803991121920} \gamma_E-\frac{54011573119031}{8269622968320} \pi^2- \frac{7132682314667557}{16883813560320} \ln 2 \\ \nonumber
& + \frac{9025682228697}{61760634880} \ln 3 - \frac{10950620263057}{803991121920} \ln v_0\bigg) v^2 v_0^8 + \bigg(-\frac{25027782258309739265857}{131539404055707648000}   \\ \nonumber
& + \frac{27973461272447}{140826470400} \gamma_E - \frac{622121304800869}{5530639564800} \pi^2 + \frac{776046310518967}{1267438233600} \ln 2 + \frac{263064734543}{1545420800} \ln 3 \\ 
& - \frac{15341796875}{100590336} \ln 5  + \frac{27973461272447}{140826470400} \ln v_0 \bigg) v_0^{10} \bigg] +\cdots+\mathcal{O}(e_0^{10}) \,,
\end{align}
\end{subequations}

where $\phi_c$ represents a reference phase (say the phase at the time of coalescence).\\ 

For $t(v)$ we get,
 
\end{widetext}

\begin{widetext}
\label{eq:hybridtime}

\begin{align}
\label{eq:timeHybrid_circ_ecc}
     t_c - t^{\rm Hybrid}(v, v_0, e_0) &= t_{\mathrm{circ}}^{\rm Hybrid}(v, v_0, e_0) + t_{\mathrm{ecc}}^{\rm Hybrid}(v, v_0, e_0)
     \,,\end{align}
where
\begin{align}
\label{eq:timeHybrid_circ}
t_{\mathrm{circ}}^{\rm Hybrid}(v, v_0, e_0) &= t^{\rm Newt}_{\rm circ}+\cdots+t^{\rm 4PN}_{\rm circ}+t^{\rm 4.5PN}_{\rm circ}+t^{\rm 5PN}_{\rm circ} \,,
\end{align}
with
\begin{subequations}
\label{eq:timeHybrid_circ_PNCoeff}
\begin{align} 
    t^{\rm Newt}_{\rm circ} &= \frac{5}{256}\frac{m}{\eta}\frac{1}{v^8} \,, \\ \nonumber
    t_{\mathrm{circ}}^{4\mathrm{PN}} &= t^{\rm Newt}_{\rm circ} \bigg[\frac{2500489942240134433}{461347517030400}-\frac{36812}{105} \gamma_E-\frac{18098}{63} \pi^2 + \bigg(-\frac{722352859826557}{70413235200}-\frac{3911888}{2205} \gamma_E \\ \nonumber
    &  +\frac{65577}{224} \pi^2 - \frac{9964112}{2205} \ln 2 +\frac{47385}{49} \ln 3 -\frac{1955944}{2205} \ln v \bigg)\eta + \bigg(\frac{1502014727}{1016064}-\frac{2255 \pi ^2}{48}\bigg) \eta ^2+\frac{258479}{4032} \eta^3 \\
    & -\frac{1195}{32} \eta^4-\frac{202204}{441}\ln 2 -\frac{47385}{196} \ln 3 -\frac{18406}{105} \ln v \bigg] v^8 \ln v\,, \\ \nonumber
t_{\mathrm{circ}}^{4.5\mathrm{PN}} &= t^{\rm Newt}_{\rm circ}\bigg[-\frac{102282756713483}{23471078400}+\frac{54784}{105} \gamma_E+\frac{512}{3} \pi^2 + \bigg(\frac{298583452147}{33530112}-\frac{902}{3} \pi^2\bigg)\eta -\frac{9058667}{31752}\eta^2  \\ 
& -\frac{2064751}{49896} \eta ^3 + \frac{109568}{105} \ln 2 +\frac{54784}{105} \ln v \bigg]\pi v^9\,, \\ \nonumber
t_{\mathrm{circ}}^{5\mathrm{PN}} &= t^{\rm Newt}_{\rm circ} \bigg[\frac{53039887927351709773}{4678025851699200}-\frac{6470582647}{27505170} \gamma_E - \frac{578223115}{3048192} \pi^2 -\frac{53992839431}{55010340} \ln 2 +\frac{5512455}{21952} \ln 3 \\ 
& - \frac{6470582647 }{27505170}\ln v \bigg]v^{10} \,.
\end{align}
\end{subequations}
And
\begin{align}
\label{eq:hybrid_time_ecc}
    t^{\rm Hybrid}_{\rm ecc}
    &= t^{\rm Newt}_{\rm ecc}+\cdots+t^{\rm 3.5PN}_{\rm ecc}+t^{\rm 4PN}_{\rm ecc}
    +t^{\rm 4.5PN}_{\rm ecc}+t^{\rm 5PN}_{\rm ecc} \,,
\end{align}
with
\begin{subequations}
\label{eq:hybrid_time_ecc_PNCoeff}
\begin{align}
    t^{\rm Newt}_{\rm ecc} &= t^{\rm Newt}_{\rm circ}\times \bigg(-\frac{157}{43}\bigg)e_0^2 \bigg(\frac{v_0}{v}\bigg)^{19/3} +\cdots+\mathcal{O}(e_0^{10}) \,,\\ \nonumber
    t^{\rm 3.5PN}_{\rm ecc} &= t^{\rm Newt}_{\rm circ}\times\bigg(-\frac{157}{43}\bigg)e_0^2 \bigg(\frac{v_0}{v}\bigg)^{19/3}\bigg[\frac{1669949401791959}{26531707023360} \pi v^{7}
- \frac{471859927369}{12562361280} \pi v^{5} v_0^{2}+ \frac{360094505303}{21799176192} \pi v^{4} v_0^{3} \\ 
& + \frac{3363921338873}{1171529920512} \pi v^{3} v_0^{4} + \frac{13456336501439}{531208419840} \pi v^{2} v_0^{5} - \frac{200064064513}{7681443840} \pi v_0^{7} \bigg] +\cdots+\mathcal{O}(e_0^{10}) \,, \\ \nonumber
t^{\rm 4PN}_{\rm ecc} &= t^{\rm Newt}_{\rm circ}\times \bigg(-\frac{157}{43}\bigg)e_0^2 \bigg(\frac{v_0}{v}\bigg)^{19/3}\bigg[\bigg( -\frac{80818373886451013723}{272215314059673600} - \frac{28385633}{508032} \gamma_E - \frac{64641510119}{2435559840} \pi^{2} \\ \nonumber
& 
- \frac{6102681415273}{22731891840} \ln 2 - \frac{2983418217}{39388160} \ln 3 - \frac{28385633}{508032} \ln v \bigg) v^{8} + \bigg( -\frac{7378818324005587696063}{16714975424716800000} \\ \nonumber
& 
+ \frac{89456686279}{2492532000} \gamma_E + \frac{2620108999447}{41019955200} \pi^{2} + \frac{1177248948661}{7477596000} \ln 2 - \frac{20764170369}{246176000} \ln 3 + \frac{89456686279}{2492532000} \ln v \bigg)v^{6} v_0^{2} \\ \nonumber
& -\frac{62792514161}{897311520} \pi^{2} v^{5} v_0^{3} - \frac{1139743046544589}{922889923264512} v^{4} v_0^{4}
- \frac{126840801139}{2050997760} \pi^{2} v^{3} v_0^{5}+ \bigg(
\frac{466768271416848150829}{989526545143234560} \\ \nonumber
& - \frac{58355048923}{737789472} \gamma_E + \frac{2160966452927}{60709533696} \pi^{2}
- \frac{212713565429}{11066842080} \ln 2
- \frac{50825365191}{364340480} \ln 3
- \frac{58355048923}{737789472} \ln v_0 \bigg)v^{2} v_0^{6}  \\ \nonumber
& +  \bigg(\frac{916726039513098761}{2372644373299200} - \frac{1574437}{362880} \gamma_E
- \frac{7765571}{3732480} \pi^{2} - \frac{1025508937}{7620480} \ln 2 + \frac{11679093}{250880} \ln 3 \\
& - \frac{1574437}{362880} \ln v_0 \bigg)v_0^{8} \bigg] +\cdots+\mathcal{O}(e_0^{10}) \,, \\ \nonumber
t^{\rm 4.5PN}_{\rm ecc} &= t^{\rm Newt}_{\rm circ}\times \bigg(-\frac{157}{43}\bigg)e_0^2 \bigg(\frac{v_0}{v}\bigg)^{19/3}\bigg[\bigg(
     \frac{25837139841614323949}{53487921359093760} + \frac{350517983}{3255552} \gamma_E
     + \frac{68314126789}{5625593856} \pi^{2} \\ \nonumber
     & - \frac{162649918793}{341832960} \ln 2 + \frac{7707912669}{11253760} \ln 3
     + \frac{350517983}{3255552} \ln v \bigg) \pi v^{9} + \frac{4730966655276619847}{26743960679546880} \pi v^{7} v_0^{2} \\ \nonumber
     & + \bigg(-\frac{981932406689059852247}{1193926816051200000} + \frac{11904401951}{178038000} \gamma_E + \frac{348669640943}{2929996800} \pi^{2} + \frac{156661790909}{534114000} \ln 2
     - \frac{2763181161}{17584000} \ln 3 \\ \nonumber
     & + \frac{11904401951}{178038000} \ln v
      \bigg)\pi v^{6} v_0^{3} + \frac{198745969005643}{37988580510720} \pi v^{5} v_0^{4} + \frac{730582083052159}{27466962001920} \pi v^{4} v_0^{5} \\ \nonumber
    &  + \bigg(-\frac{74796691155101107993}{64949618793185280} + \frac{9351030991}{48426336} \gamma_E - \frac{346281335459}{3984795648} \pi^{2}
    + \frac{34086016193}{726395040} \ln 2
    + \frac{8144446347}{23914240} \ln 3  \\ \nonumber
    & + \frac{9351030991}{48426336} \ln v_0 \bigg) \pi v^{3} v_0^{6} - \frac{3519670868975080847}{44978479324692480} \pi v^{2} v_0^{7}+ \bigg(
    \frac{5549358074328361}{5323240581120}
    - \frac{1556101}{9072} \gamma_E
    + \frac{22026761}{2239488} \pi^{2} \\
    &  - \frac{5269643}{136080} \ln 2 - \frac{1366497}{4480} \ln 3 - \frac{1556101}{9072} \ln v_0 \bigg)\pi v_0^{9} \bigg]+\cdots+\mathcal{O}(e_0^{10}) \,, \\ \nonumber
    t^{\rm 5PN}_{\rm ecc} &= t^{\rm Newt}_{\rm circ}\times \bigg(-\frac{157}{43}\bigg)e_0^2 \bigg(\frac{v_0}{v}\bigg)^{19/3}\bigg[\bigg( \frac{776108917708285695087018041}{134235961838849654784000}
    - \frac{22705732129814659}{17597560780800} \gamma_E \\ \nonumber
    & 
    - \frac{11897104210386999449}{11288035351756800} \pi^2 + \frac{1357681571333591809}{369548776396800} \ln 2
    - \frac{1065001883374883}{286745804800} \ln 3
    - \frac{292245888671875}{205304875776} \ln 5  \\ \nonumber
    & - \frac{22705732129814659}{17597560780800} \ln v
    \bigg) v^{10} + \bigg(
    - \frac{228958453220315721877259}{274393036572150988800}
    - \frac{80416498289}{512096256} \gamma_E \\ \nonumber
    & - \frac{183129398167127}{2455044318720} \pi^2 - \frac{17288896449468409}{22913746974720} \ln 2
    - \frac{939113756529}{4411473920} \ln 3
    - \frac{80416498289}{512096256} \ln v \bigg) v^8 v_0^2 \\ \nonumber
    & + \frac{629570924475568543}{1910282905681920} \pi^2  v^7 v_0^3 + \bigg(
    \frac{3107935878552061956906061}{50546085684343603200000}
    - \frac{37678884701413}{7537416768000} \gamma_E \\ \nonumber
    & - \frac{1103581956829909}{124044344524800} \pi^2
    - \frac{495853683458767}{22612250304000} \ln 2 + \frac{323918723209}{27571712000} \ln 3
    - \frac{37678884701413}{7537416768000} \ln v \bigg) v^6 v_0^4  \\ \nonumber
    & 
    - \frac{127397350196233}{1130612515200} \pi^2
     v^5 v_0^5 + \bigg(
    \frac{25342152821305345008749}{51165017345784545280} - \frac{102201888773}{1230598656} \gamma_E
    + \frac{117324902837887}{3139081371648} \pi^2  \\ \nonumber
    & - \frac{11548813431349}{572228375040} \ln 2 - \frac{102201888773}{697733120} \ln 3 - \frac{102201888773}{1230598656} \ln v_0 \bigg) v^4 v_0^6 + \frac{564005205742082099}{2952255399690240} \pi^2  v^3 v_0^7
    \\ \nonumber
    & + \bigg(
    \frac{16127703613136843321521159}{13892952693811013222400} - \frac{27698627724203}{2124833679360} \gamma_E
    - \frac{136617508477549}{21855432130560} \pi^2  \\ \nonumber
    & - \frac{18041490560629703}{44621507266560} \ln 2 + \frac{22829666813763}{163224535040} \ln 3 - \frac{27698627724203}{2124833679360} \ln v_0
    \bigg) v^2 v_0^8
     \\ \nonumber
    & + \bigg(- \frac{25027782258309739265857}{131539404055707648000} + \frac{27973461272447}{140826470400} \gamma_E
    - \frac{622121304800869}{5530639564800} \pi^2 + \frac{776046310518967}{1267438233600} \ln 2 \\ 
    & + \frac{263064734543}{1545420800} \ln 3
    - \frac{15341796875}{100590336} \ln 5 + \frac{27973461272447}{140826470400} \ln v_0\bigg) v_0^{10} \bigg] +\cdots+\mathcal{O}(e_0^{10}) \,,
\end{align}
\end{subequations}
\end{widetext}
\noindent
where $t_c$ is time of coalescence. Note that our results include the BH horizon contributions available at orders 4PN and beyond in the test particle limit. Special care must be taken while comparing the circular limit of our results with that of Ref.~\cite{Blanchet:2023bwj} which does not compute these contributions. We discuss the details of such comparisons in Appendix~\ref{sec:BHhorizonContributions}.

\vspace{0pt}
%%%%%%%%%%%%%%%%%
%%%%%%%%%%%%%%%%%

\subsection{\TFtwo{} hybrid phase}
\label{sec:hybTF2phase}
In this section we present a frequency-domain version of the \TTtwo{} approximant, referred to as \TFtwo{}. As discussed earlier, this too is also fully analytical and is obtained under SPA~\cite{bender1999, Damour:2000gg}. Following the footsteps of Ref.~\cite{Moore:2016qxz}, which provides 3PN version of this approximant at the leading order in eccentricity, we provide a 5PN version of the hybrid phase with eccentricity corrections up to $\mathcal{O}(e_{0}^{10})$ and refer to Ref.~\cite{Moore:2016qxz} for details. It reads, 

\begin{widetext}
\label{eq:AppSPAphase}
    
\begin{align}
\label{eq:SPAphaseHybrid_circ_ecc}
     \Psi^{\rm Hybrid}(v, v_0, e_0) &= \psi_0 + 2\pi f t_0 + \Psi_{\mathrm{circ}}^{\rm Hybrid}(v, v_0, e_0) + \Psi_{\mathrm{ecc}}^{\rm Hybrid}(v, v_0, e_0)
     \,,\end{align}
where
\begin{align}
\label{eq:SPAphaseHybrid_circ}
\Psi_{\mathrm{circ}}^{\rm Hybrid}(v, v_0, e_0) &= \Psi^{\rm Newt}_{\rm circ}+\cdots+\Psi^{\rm 4PN}_{\rm circ}+\Psi^{\rm 4.5PN}_{\rm circ}+\Psi^{\rm 5PN}_{\rm circ} \,,
\end{align}
with
\begin{subequations}
\label{eq:SPAphaseHybrid_circ_PNCoeff}
\begin{align}
\Psi^{\rm Newt}_{\rm circ} &= \frac{3}{128}\frac{1}{\eta v^5} \,, \\ \nonumber
    \Psi_{\mathrm {circ}}^{4\mathrm{PN}} &= \Psi^{\rm Newt}_{\rm circ}\bigg[-\frac{2554404624135128353}{276808510218240} +\frac{36812}{63} \gamma_E+\frac{90490}{189} \pi^2  + \bigg(\frac{680712846248317}{42247941120}-\frac{109295}{224} \pi^2+\frac{3911888}{1323} \gamma_E  \\ \nonumber
    & + \frac{9964112}{1323} \ln 2 -\frac{78975}{49} \ln 3 + \frac{1955944}{1323} \ln v \bigg)\eta +\bigg(-\frac{7510073635}{3048192} + \frac{11275}{144} \pi^2 \bigg) \eta ^2  -\frac{1292395}{12096} \eta^3 + \frac{5975 }{96}\eta^4  \\ 
    & +\frac{1011020 }{1323}\ln 2 +\frac{78975}{196} \ln 3 + \frac{18406}{63}\ln v \bigg]v^8 \ln v \,, \\ \nonumber
    \Psi_{\mathrm {circ}}^{4.5\mathrm{PN}} &= \Psi^{\rm Newt}_{\rm circ}\bigg[\frac{105344279473163}{18776862720} 
    -\frac{13696}{21} \gamma_E -\frac{640}{3} \pi^2 + \bigg(-\frac{1492917260735}{134120448}+\frac{2255}{6} \pi^2 \bigg)\eta  +\frac{45293335}{127008} \eta ^2 \\ 
    & +\frac{10323755 }{199584}\eta^3 -\frac{27392}{21} \ln 2 - \frac{13696}{21}\ln v \bigg]\pi v^9 \,, \\ \nonumber
    \Psi_{\mathrm {circ}}^{5\mathrm{PN}} &= \Psi^{\rm Newt}_{\rm circ}\bigg[-\frac{1438019696284910204959}{126306697995878400} +\frac{6470582647}{27505170} \gamma_E + \frac{578223115}{3048192} \pi^2 + \frac{53992839431}{55010340} \ln 2 - \frac{5512455}{21952} \ln
    3 \\
    & +\frac{6470582647}{27505170} \ln v \bigg] v^{10} \,. \\ \nonumber
\end{align}

\end{subequations}
And
\begin{align}
\label{eq:hybrid_SPAPhase_ecc}
    \Psi^{\rm Hybrid}_{\rm ecc}
    &= \Psi^{\rm Newt}_{\rm ecc}+\cdots+\Psi^{\rm 3.5PN}_{\rm ecc}+\Psi^{\rm 4PN}_{\rm ecc}
    +\Psi^{\rm 4.5PN}_{\rm ecc}+\Psi^{\rm 5PN}_{\rm ecc} \,,
\end{align}
where
\begin{subequations}
\label{eq:hybrid_SPAPhase_ecc_PNCoeff}
\begin{align} 
    \Psi^{\rm Newt}_{\rm ecc} &= \Psi^{\rm Newt}_{\rm circ}\times\bigg(-\frac{2355}{1462}\bigg)e_0^2\bigg(\frac{v_0}{v}\bigg)^{19/3}+\cdots+ \mathcal{O}(e_0^{10})\\ \nonumber
    \Psi_{\rm ecc}^{3.5\mathrm{PN}} &= \Psi^{\rm Newt}_{\rm circ}\times\bigg(-\frac{2355}{1462}\bigg)e_0^2\bigg(\frac{v_0}{v}\bigg)^{19/3}\bigg[\frac{28389139830463303}{172456095651840} \pi v^7 - \frac{8021618765273}{119342432160} \pi v^5 v_0^2
    + \frac{6121606590151}{239790938112} \pi v^4 v_0^3 \\
    & + \frac{3363921338873}{861419059200} \pi v^3 v_0^4 + \frac{228757720524463}{7436917877760} \pi v^2 v_0^5- \frac{200064064513}{7681443840} \pi v_0^7 \bigg] +\cdots+ \mathcal{O}(e_0^{10}) \,, \\ \nonumber
    \Psi_{\rm ecc}^{4\mathrm{PN}} &= \Psi^{\rm Newt}_{\rm circ}\times\bigg(-\frac{2355}{1462}\bigg)e_0^2\bigg(\frac{v_0}{v}\bigg)^{19/3}\bigg[\bigg(- \frac{1451481722585114957131}{1361076570298368000}
    - \frac{482555761}{2540160} \gamma_E
    - \frac{1098905672023}{12177799200} \pi^2 \\ \nonumber
    & - \frac{103745584059641}{113659459200} \ln 2 - \frac{50718109689}{196940800} \ln 3 - \frac{482555761}{2540160} \ln v \bigg) v^8 +
    \bigg(
    - \frac{123527734905416610887471}{133719803397734400000} \\ \nonumber
    & 
     + \frac{1520763666743}{19940256000} \gamma_E + \frac{44541852990599}{328159641600} \pi^2 + \frac{20013232127237}{59820768000} \ln 2
    - \frac{352990896273}{1969408000} \ln 3 \\ \nonumber
    & + \frac{1520763666743}{19940256000} \ln v
    \bigg) v^6 v_0^2 - \frac{1067472740737}{8524459440} \pi^2 v^5 v_0^3 - \frac{19375631791258013}{10151789155909632}
    v^4 v_0^4 
    - \frac{2156293619363}{25637472000} \pi^2
    v^3 v_0^5  \\ \nonumber
    & +\bigg(
    \frac{7935060614086418564093}{13853371632005283840}
    - \frac{992035831691}{10329052608} \gamma_E + \frac{36736429699759}{849933471744} \pi^2
    - \frac{3616130612293}{154935789120} \ln 2 \\ \nonumber
    &  - \frac{864031208247}{5100766720} \ln 3
    - \frac{992035831691}{10329052608} \ln v_0
    \bigg) v^2 v_0^6 + \bigg(
    \frac{916726039513098761}{2372644373299200}
    - \frac{1574437}{362880} \gamma_E
    - \frac{7765571}{3732480} \pi^2
    \\ 
     & - \frac{1025508937}{7620480} \ln 2
    + \frac{11679093}{250880} \ln 3
    - \frac{1574437}{362880} \ln v_0 \bigg) v_0^8 \bigg] +\cdots+ \mathcal{O}(e_0^{10}) \,, \\ \nonumber
    \Psi_{\rm ecc}^{4.5\mathrm{PN}} &= \Psi^{\rm Newt}_{\rm circ}\times\bigg(-\frac{2355}{1462}\bigg)e_0^2\bigg(\frac{v_0}{v}\bigg)^{19/3}\bigg[\bigg(
        \frac{481189256111418129853}{187207724756828160} 
        + \frac{5958805711}{11394432} \gamma_E + \frac{1161340155413}{19689578496} \pi^2 \\ \nonumber
        & 
        - \frac{2765048619481}{1196415360} \ln 2 + \frac{131034515373}{39388160} \ln 3 + \frac{5958805711}{11394432} \ln v
        \bigg)\pi v^{9} + \frac{80426433139702537399}{173835744417054720} \pi
         v^{7} v_{0}^{2} \\ \nonumber
        & + \bigg(
        - \frac{16438389007886361561799}{9551414528409600000}
        + \frac{202374833167}{1424304000} \gamma_E + \frac{5927383896031}{23439974400} \pi^{2} + \frac{2663250445453}{4272912000} \ln 2 \\ \nonumber 
        & 
        - \frac{46974079737}{140672000} \ln 3
        + \frac{202374833167}{1424304000} \ln v
        \bigg)\pi v^{6} v_{0}^{3} + \frac{3378681473095931}{360891514851840} \pi
        v^{5} v_{0}^{4} +
        \frac{12419895411886703}{302136582021120} \pi
        v^{4} v_{0}^{5} \\ \nonumber
        & + \bigg(
        - \frac{74796691155101107993}{47757072642048000}
        + \frac{9351030991}{35607600} \gamma_E - \frac{346281335459}{2929996800} \pi^{2} + \frac{34086016193}{534114000} \ln 2 \\ \nonumber 
        & 
        + \frac{8144446347}{17584000} \ln 3 + \frac{9351030991}{35607600} \ln v_{0}
        \bigg)\pi v^{3} v_{0}^{6} - \frac{59834404772576374399}{629698710545694720} \pi
        v^{2} v_{0}^{7} \\ \nonumber
        & +\bigg(
        \frac{5549358074328361}{5323240581120}
        - \frac{1556101}{9072} \gamma_E
        + \frac{22026761}{2239488} \pi^{2}
        - \frac{5269643}{136080} \ln 2 - \frac{1366497}{4480} \ln 3
        - \frac{1556101}{9072} \ln v_{0} \bigg)\pi v_{0}^{9} \bigg] \\ 
        & +\cdots+ \mathcal{O}(e_0^{10}) \,, \\ \nonumber
        \Psi_{\rm ecc}^{5\mathrm{PN}} &= \Psi^{\rm Newt}_{\rm circ}\times\bigg(-\frac{2355}{1462}\bigg)e_0^2\bigg(\frac{v_0}{v}\bigg)^{19/3}\bigg[\bigg(
        \frac{10985531122523926377960368617}{268471923677699309568000}
        - \frac{385997446206849203}{35195121561600} \gamma_E \\ \nonumber
        & - \frac{202250771576578990633}{22576070703513600} \pi^2 + \frac{23080586712671060753}{739097552793600} \ln 2
        - \frac{18105032017373011}{573491609600} \ln 3 \\ \nonumber
        & - \frac{4968180107421875}{410609751552} \ln 5 - \frac{385997446206849203}{35195121561600} \ln v
        \bigg) v^{10} + \bigg(
        - \frac{4112047720083630673552123}{1371965182860754944000} \\ \nonumber
        & - \frac{1367080470913}{2560481280} \gamma_E - \frac{3113199768841159}{12275221593600} \pi^2 - \frac{293911239640962953}{114568734873600} \ln 2
        - \frac{15964933860993}{22057369600} \ln 3  \\ \nonumber
        & - \frac{1367080470913}{2560481280} \ln v
        \bigg) v^{8} v_0^{2} +
        \frac{823285055083435787}{955141452840960} \pi^2
        v^{7} v_0^{3} + \bigg(
        \frac{52029506955038219681639837}{404368685474748825600000}
        \\ \nonumber
        & - \frac{640541039924021}{60299334144000} \gamma_E - \frac{18760893266108453}{992354756198400} \pi^2
        - \frac{8429512618799039}{180898002432000} \ln 2 + \frac{5506618294553}{220573696000} \ln 3
         \\ \nonumber
        & - \frac{640541039924021}{60299334144000} \ln v
        \bigg) v^{6} v_0^{4} - \frac{2165754953335961}{10740818894400} \pi^2
        v^{5} v_0^{5} + \bigg(
        \frac{430816597962190865148733}{562815190803629998080} \\ \nonumber
        & - \frac{1737432109141}{13536585216} \gamma_E
        + \frac{1994523348244079}{34529895088128} \pi^2 - \frac{196329828332933}{6294512125440} \ln 2 - \frac{1737432109141}{7675064320} \ln 3 \\ \nonumber
        &  - \frac{1737432109141}{13536585216} \ln v_0
        \bigg) v^{4} v_0^{6} +
        \frac{564005205742082099}{2170776029184000} \pi^2
        v^{3} v_0^{7} + \bigg(
         \frac{274170961423326336465859703}{194501337713354185113600} \\ \nonumber
        & - \frac{470876671311451}{29747671511040} \gamma_E
        - \frac{2322497644118333}{305976049827840} \pi^2 - \frac{306705339530704951}{624701101731840} \ln 2 + \frac{388104335833971}{2285143490560} \ln 3  \\ \nonumber
        & - \frac{470876671311451}{29747671511040} \ln v_0
        \bigg) v^{2} v_0^{8} + \bigg(
        - \frac{25027782258309739265857}{131539404055707648000}
        + \frac{27973461272447}{140826470400} \gamma_E \\ \nonumber
        & - \frac{622121304800869}{5530639564800} \pi^2
        + \frac{776046310518967}{1267438233600} \ln 2 + \frac{263064734543}{1545420800} \ln 3
        - \frac{15341796875}{100590336} \ln 5 \\ 
        & 
        + \frac{27973461272447}{140826470400} \ln v_0 \bigg) v_0^{10}\bigg]+\cdots+ \mathcal{O}(e_0^{10}) \,.
\end{align}

\end{subequations}
\end{widetext}

Note that some of the contributions at 2.5PN and 4PN order beyond the leading term have been absorbed into a redefinition of the constant phase ($\psi_0$) and the linear phase term involving $t_0$ to simplify the output following \cite{Blanchet:2023bwj}. Since the 2.5PN term is completely degenerate with the initial phase, and the 4PN term with linear initial phase, such absorption has no consequence for data analyses applications. 

%%%%%%%%%%%%%%%%%
%%%%%%%%%%%%%%%%%

\subsection{The mode amplitude}
\label{sec:hyb_22_mode_amp}
Following the prescription to compare the mode amplitudes from the PN and BHP approaches presented in Sec.~\ref{sec:modes} we now express the final hybrid 22-mode amplitude ($\hat{H}^{22}$) purely in terms of the PN parameters through $\mathcal{O}(v^{10},e_t^{10})$ which takes the following form\footnote{For the hybrid amplitudes, we replace $Z^{\infty}_{\ell mn}$ in Eq.~\eqref{eq:Hlm_bhp_def} with $(-1)^n\,Z^{\infty}_{\ell mn}$ since in Ref.~\cite{Fujita:2020zxe} the initial position of radial motion is apastron, while in PN it is periastron. See Ref.~\cite{Drasco:2005is} to account for the overall factor $(-1)^n$ in $Z^{\infty}_{\ell mn}$ by transforming the initial position of radial motion from apastron to periastron.} \\
    
\begin{widetext}
\label{eq:H22Hyb}

\begin{align}
\label{eq:H22Hyb_circ_ecc}
     \hat{H}_{\rm Hybrid}^{22}(v, e_t, \xi) &=   \hat{H}_{\rm Hybrid}^{\rm circ}(v, e_t, \xi) + \hat{H}_{\rm Hybrid}^{\rm ecc}(v, e_t, \xi)
     \,,\end{align}
     
where the circular hybrid 22-mode amplitude reads
\begin{align}
\label{eq:H22Hybrid_circ}
\hat{H}^{\mathrm{circ}}_{\rm Hybrid}(v, e_t, \xi) &= 1+\cdots+ \hat{H}^{\rm 3.5PN}_{\rm circ}+ \hat{H}^{\rm 4PN}_{\rm circ}+ \hat{H}^{\rm 4.5PN}_{\rm circ} + \hat{H}^{\rm 5PN}_{\rm circ} \,,
\end{align}

with
\begin{subequations}
\label{eq:H22Hybrid_circ_PNCoeff}
\begin{align}
\hat{H}^{3.5\mathrm{PN}}_{\mathrm{circ}} &= \bigg[-\frac{2173}{756} \pi + \bigg(-\frac{2495}{378}\pi +\frac{14333}{162} i\bigg)\eta+\bigg(\frac{40}{27} \pi -\frac{4066}{945}i\bigg)\eta^2 \bigg] v^7 \,, \\
\hat{H}^{4\mathrm{PN}}_{\mathrm{circ}} &= \bigg[-\frac{846557506853}{12713500800}+\frac{45796}{2205}\gamma_E-\frac{22898}{2205} i \pi-\frac{107}{63}\pi^2+  \bigg(-\frac{336005827477}{4237833600}+\frac{15284}{441} \gamma_E-\frac{219314}{2205}i\pi  \nonumber \\
& -\frac{9755}{32256}\pi^2 + \frac{30568}{441} \ln 2+\frac{15284}{441} \ln v \bigg)\eta+\bigg(\frac{256450291}{7413120}-\frac{1025}{1008} \pi ^2\bigg)\eta^2-\frac{81579187}{15567552} \eta^3+\frac{26251249}{31135104}\eta^4 \nonumber \\
& +\frac{91592}{2205} \ln 2 +\frac{45796}{2205} \ln v \bigg]v^8 \,, \\
\hat{H}^{4.5\mathrm{PN}}_{\mathrm{circ}} &= \bigg[-\frac{259}{81}i+ \frac{27027409}{323400}\pi-\frac{1712}{105}\pi \gamma_E+\frac{856}{63}i\pi^{2}-\frac{4}{3}\pi^{3} -\frac{3424}{105}\pi \ln 2-\frac{1712}{105}\pi \ln v-\frac{64}{3}i \zeta(3) \bigg]v^{9} \,, \\ 
\hat{H}^{5\mathrm{PN}}_{\mathrm{circ}} &= \bigg[-\frac{866305477369}{9153720576}+\frac{232511}{19845}\gamma_E-\frac{232511}{39690}i\pi-\frac{2173}{2268}\pi^{2}+\frac{465022}{19845}\ln 2 + \frac{232511}{19845}\ln v \bigg]v^{10} \,.
\end{align}
\end{subequations}

Whereas, the eccentric hybrid 22-mode amplitude reads
\begin{align}
\label{eq:H22Hybrid_ecc}
\hat{H}^{\mathrm{ecc}}_{\rm Hybrid}(v, e_t, \xi) &= \hat{H}^{\rm Newt}_{\rm ecc}+\cdots+ \hat{H}^{\rm 3.5PN}_{\rm ecc}+ \hat{H}^{\rm 4PN}_{\rm ecc}+ \hat{H}^{\rm 4.5PN}_{\rm ecc} + \hat{H}^{\rm 5PN}_{\rm ecc} \,,
\end{align}

where
\begin{subequations}
\label{eq:H22Hybrid_ecc_PNCoeff}
\begin{align}
\hat{H}^{\rm Newt}_{\rm ecc} &= e_t\bigg[\frac{1}{4}\mathrm{e}^{-i\xi}+ \frac{5}{4}\mathrm{e}^{i\xi}  \bigg] + \cdots+\mathcal{O}(e_t^{10}) \,, \\ \nonumber
\hat{H}^{3.5\mathrm{PN}}_{\rm ecc} &= e_t\bigg[
\bigg(-\frac{135}{4}i -\frac{10243}{756} \pi -\frac{135}{2} i \ln 2 + \frac{135}{2} i \ln 3 \bigg)\mathrm{e}^{-i\xi} +
\bigg(-\frac{1419}{28}i -\frac{5191}{756} \pi +\frac{40949}{378} i \ln 2 \bigg)\mathrm{e}^{i\xi}
\bigg]v^7  \\ 
& + \cdots+\mathcal{O}(e_t^{10}) \,, \\ \nonumber
\hat{H}^{4\mathrm{PN}}_{\rm ecc} &=  e_t\bigg\{
\bigg[-\frac{81636771787}{635675040} +\frac{534679}{8820} \gamma_E -\frac{1249099}{17640}i \pi  - \frac{4997}{1008}\pi^2  + \bigg(\frac{172931}{8820} + \frac{81}{4}i \pi  -\frac{81}{2} \ln 3 \bigg)\ln 2  \\ \nonumber
& + \frac{81}{4} \ln^2 2 + \bigg(\frac{14229}{140} - \frac{81}{4}i \pi \bigg)\ln 3 +\frac{81}{4} \ln^2 3 +\frac{534679}{8820} \ln v   \bigg]\mathrm{e}^{-i\xi}  + \bigg[\frac{647217469}{5556600}-\frac{160393}{8820}\gamma_E \\
& + \frac{81013}{17640}i \pi + \frac{1499}{1008} \pi^2 +\bigg(-\frac{16927}{252} + \frac{599}{28}i \pi \bigg)\ln 2 +\frac{599}{28} \ln^2 2  -\frac{160393}{8820} \ln v \bigg]\mathrm{e}^{i\xi} \bigg\}v^8 + \cdots+\mathcal{O}(e_t^{10}) \,, \\ \nonumber
\hat{H}^{4.5\mathrm{PN}}_{\rm ecc} &= e_t\bigg\{\bigg[-\frac{477929}{2592}i + \frac{6119050423}{11642400}\pi-\frac{19153}{210}\pi\gamma_E + \frac{19153}{252}i\pi^2 -\frac{179}{24}\pi^3 +\bigg(-\frac{423883773}{215600}i +\frac{8667}{35}i\gamma_E  \\ \nonumber
&  + \frac{6848}{105}\pi - \frac{81}{4}i\pi^2 + 243\pi \ln 3+243i\ln^2 3 \bigg)\ln 2 + \bigg( \frac{8667}{35}i-\frac{243}{2}\pi - 243 i  \ln 3 \bigg)\ln^2 2 + 81 i\ln^3 2 \\ \nonumber
& +\bigg(\frac{423883773}{215600}i -\frac{8667}{35}i\gamma_E - \frac{8667}{35}\pi  +\frac{81}{4}i\pi^2 \bigg)\ln 3 -\bigg(\frac{8667}{35}i+\frac{243}{2}\pi \bigg) \ln^2 3  - 81 i \ln ^3 3  \\ \nonumber
& + \bigg(- \frac{19153}{210}\pi +\frac{8667}{35}i \ln 2-\frac{8667}{35}i \ln 3 \bigg)\ln v -\frac{358}{3}i\zeta (3) \bigg]\mathrm{e}^{-i\xi}+\bigg[-\frac{6252193}{18144}i+ \frac{189709913}{1293600}\pi \\ \nonumber
& -\frac{6527}{210} \pi \gamma_E + \frac{6527}{252}i\pi^2 -\frac{61}{24}\pi^3 +  \bigg(\frac{3618422231}{5821200}i-\frac{107}{35}i\gamma_E - \frac{6848}{105}\pi + \frac{1}{4}i\pi^2 \bigg)\ln 2 + \bigg(-\frac{107}{35}i \\ 
& +\frac{3}{2}\pi \bigg)\ln ^2 2   -i \ln^3 2 - \bigg(\frac{6527}{210} \pi + \frac{107}{35}i \ln 2 \bigg) \ln v -\frac{122}{3}i \zeta (3)   \bigg]\mathrm{e}^{i\xi} \bigg\}v^9 + \cdots+\mathcal{O}(e_t^{10}) \,, \\ \nonumber
\hat{H}^{5\mathrm{PN}}_{\rm ecc} &=  e_t\bigg\{\bigg[-\frac{4543208867989}{10401955200}+\frac{2738023}{39690}\gamma_E -\frac{3780179}{39690}i\pi - \frac{25589}{4536}\pi^2 + \bigg(\frac{556790}{3969} - \frac{243}{2} i\pi + 243 \ln 3 \bigg)\ln 2    \\ \nonumber
& -\frac{243}{2}\ln^2 2 + \bigg(-\frac{81}{35}+\frac{243}{2} i\pi \bigg)\ln 3 -\frac{243}{2} \ln^2 3+\frac{2738023}{39690} \ln v \bigg]\mathrm{e}^{-i\xi}+\bigg[\frac{11717415191089}{38140502400}-\frac{315971}{13230}\gamma_E \\
& - \frac{691097}{13230}i\pi +\frac{2953}{1512}\pi^2 + \bigg(\frac{8951}{54} i\pi  - \frac{1369462}{3969} \bigg)\ln 2 +\frac{8951}{54}\ln^2 2-\frac{315971}{13230} \ln v  \bigg]\mathrm{e}^{i\xi} \bigg\}v^{10} + \cdots+\mathcal{O}(e_t^{10}) \,,
\end{align}
\end{subequations}

\begin{table*}[h!]
\captionsetup{justification=raggedright, singlelinecheck=false}
\caption{Contributions from mass ratio ($\eta$) dependent terms in \TTtwo{} phase -- Eq.~\eqref{eq:phaseHybrid_circ_ecc} -- for two representative IMRIs with varying PN and eccentricity orders are shown. Boxes with `-' indicate that our model does not have contributions at that order. For the circular case, number of total number GW cycles at different PN orders are also included in bold letters within the parenthesis; see for instance minor column 1 ($\phi_{\rm circ}$) of major column 2 and 3. Estimates are obtained for a fixed initial eccentricity ($e_0$) of 0.3. Lower frequency cut-off is chosen as 0.01Hz and the upper frequency cut-off is taken as the frequency of the last stable circular orbit for a test particle around a Schwarzschild BH.}
% \vspace{-18pt}
\label{table:GWcycles}
\begin{center}
\resizebox{\textwidth}{!}{
\begin{tabular}{|c|c|c|c|c|c|c|c|c|c|c|c|c|c|c|c|c|c|c|}
 \hline
\hline
 \shortstack{PN order} & 
 \multicolumn{6}{|c|}{\shortstack{$\Delta N^{\rm \leq 4.5PN}_{\rm gw}$ \\ $(m_1, m_2)=(100, 10) M_\odot$ \\ $f_{\rm cut}=40 $ Hz }}  & \multicolumn{6}{|c|}{\shortstack{$\Delta N^{\rm \leq 4.5PN}_{\rm gw}$ \\ $(m_1, m_2)=(1000, 100) M_\odot$ \\ $f_{\rm cut}=4 $ Hz }} \\
\hline 
  & 
  $\phi_{\rm circ} $ & $\phi (e^2)$ & $\phi (e^4)$ & $\phi (e^6)$ & $\phi (e^8)$& $\phi (e^{10})$  & $\phi_{\rm circ} $ & $\phi (e^2)$ & $\phi (e^4)$ & $\phi (e^6)$ & $\phi (e^8)$ & $\phi (e^{10})$  \\
\hline
0PN & 
0 & 0 & 0 & 0 & 0 & 0 & 0 & 0 & 0 & 0 & 0 & 0
\\ 
\hline
1PN  & 
2.68 $\times 10^3$ $\bf{(2.87 \times 10^4)}$  & -6.59 & -67.3 & -62.3 & -62.5
& -62.5 
 & 267 $\bf{(2.87\times 10^3)}$ & -0.659 & -6.73 & -6.23 & -6.25 
 & -6.25 
 \\ 
\hline
1.5PN  & 
0 & 0 & 0 & 0 & 0 & 0  
 & 0 & 0 & 0 & 0 & 0 & 0 
 \\ 
\hline
 2PN & 
 10.4 \bf{(76.4)} 
 & -0.897 & -1.24 & -1.20 
 & -1.20 
 & -1.20 
 & 4.45 \bf{(32.7)}
 & -0.416 & -0.577 & -0.557 & -0.558 & -0.558 
 \\ 
\hline
2.5PN & 
0.702 \bf{(-59.4)}
& -0.019 & -0.028 & -0.027 
& -0.027 
& -0.027 
 & 0.507 \bf{(-42.9)}
 & -0.019 & -0.028 & -0.027 
 & -0.027 
 & -0.027  
 \\ 
\hline
3PN &  
3.14 \bf{(1.81)} 
& -0.012 & -0.014 
& -0.014 
& -0.014 
& -0.014 
 & 2.90 \bf{(2.11)}
 & -0.026 & -0.030 
 & -0.030 
 & -0.030 
 & -0.030 
 \\ 
\hline
3.5PN  & 
-0.157 {\bf{(-2.54)}}
& - & - & - & - & - 
 & -0.155 \bf{(-2.50)}
 & - & - & - & - & -   
 \\ 
\hline
4PN  & 
1.16 \bf{(-2.65)} 
& - & - & - & - & - 
 & 1.15 \bf{(-2.64)}
 & - & - & - & - & -  
 \\ 
\hline
4.5PN & 
-0.802 \bf{(2.03)}
& - & - & - & - & - 
 & -0.802 \bf{(2.03)}
 & - & - & - & - & - 
 \\ 
\hline
Total & 
2.69 $\times 10^3$ (${\bf 2.87 \times 10^4 }$) & -7.52 & -68.6 & -63.5 & -63.7 & -63.7 
 & 275 (${\bf 2.85 \times 10^3 }$)  & -1.12 & -7.37 & -6.84 & -6.86 & -6.86  
 \\ 
\hline
\hline
\end{tabular}
}
\end{center}
\end{table*}

\end{widetext}

\clearpage

\noindent
where $\zeta (n)$ is the zeta function \cite{Titchmarsh:1986}. Note that, unlike the phase, we provide mode amplitudes in terms of the parameter pair, ($v, e_t$). Primary reason for doing so is to be able to compare the expressions in the 3PN limit of ~\cite{Boetzel:2019nfw}. A second reason involved limited accuracy of a result written in terms of ($v, v_0, e_0$) compared to the one obtained by allowing evolution of $e_t$ in mode expressions given in Eq.~\eqref{eq:str_et2e0}. This, however, could not be avoided for phase which requires one to write the RHS of Eq.~\eqref{eq:dphi_dv} or Eq.~\eqref{eq:dphi_dv_TT2} purely in terms of the frequency dependent PN parameter, $v$. Note also we list here only the expression for the quadrupolar ($\hat{H}^{22}$) mode. We will provide expression for all modes (up to $\ell=|m|=12$) as part of the \textbf{supplemental material}~\cite{supplementalfile}.

% \clearpage

%%%%%%%%%%%%%%%%%%%%%%%%%%%%%%%%%%%%%%%%%%%%%%%%%%%
%%%%%%%%%%%%%%%%%%%%%%%%%%%%%%%%%%%%%%%%%%%%%%%%%%%

\section{Number of GW cycles estimates}
\label{sec:GWcycles}
We highlighted earlier in the manuscript that the two approaches complement each other beyond the fact that BHP results can be seen as a limiting case of those from PN approach. In this section, we attempt to demonstrate this using number of GW cycles estimates, obtained using the hybrid \TTtwo{} phase presented in Sec.~\ref{sec:hybTT2phase}, in the dHz band. The number of GW cycles ($N_{\rm gw}$) in a band ($f_{0}, f_{\rm cut}$) is given by~\cite{Moore:2016qxz}  
\begin{align}
\label{eq:gwcycles}
  N_{\rm gw}=\frac{\langle\phi^{\rm Hybrid}(f_{\rm cut})\rangle-\langle\phi^{\rm Hybrid}(f_{0})\rangle}{\pi}\,.  
\end{align}
We choose to work with a $f_0$ of 0.01Hz while $f_{\rm cut}$ is taken to be the GW frequency of the dominant mode at the innermost stable circular orbit of a test particle around a Schwarzschild BH, given by 
\begin{align}
f_\textrm{isco}=(6^{3/2}\pi m)^{-1}.
\end{align}

Table~\ref{table:GWcycles} displays these estimates for two representative systems, each having an intermediate and/or a  stellar mass BH -- $(100, 10)M_{\odot}$ and $(1000, 100)M_{\odot}$, while Table~\ref{table:GWcycles2} shows these for $(10000, 10)M_{\odot}$ system -- chosen to appropriately highlight the complementary nature of the results from the two approaches. Additionally, we also compare the estimates by restricting the eccentricity order. We bullet some of our key observations below,

\begin{table}[t!]
\captionsetup{justification=raggedright, singlelinecheck=false}
\caption{
Contributions from terms beyond 4.5PN(3PN) for circular(eccentric) part in \TTtwo{} phase (Eq.~\eqref{eq:phaseHybrid_circ_ecc}) for a representative IMRI with varying PN and eccentricity orders are shown. (Also included (in bold) are contributions which can also be computed using PN results; they help us demonstrate that the number of cycles estimates for the circular case do not converge up to the 5PN order.) Choices for $e_0$ as well as that for the lower and upper frequency cut-off are same as in Table~\ref{table:GWcycles}. We note that, while the $N_{\rm GW}$ estimates seem to converge at all eccentricity orders, this is not true for the estimates for the circular part; see also Table~\ref{table:12PNorderphase}. Further, we also note that $N_{\rm GW}$ estimates become $\mathcal{O}(1)$ at 5PN, indicating the adequateness of eccentricity model presented here in analyzing for systems as heavy as the one chosen here.}
% \vspace{-12pt}
\label{table:GWcycles2}
\begin{center}
\setlength{\tabcolsep}{3.8pt}
\begin{tabular}{|c|c|c|c|c|c|c|}
\hline
 \shortstack{PN order} &
 \multicolumn{6}{|c|}{\shortstack{$ N^{\rm \geq 3.5PN}_{\rm gw}$ \\ $(m_1, m_2)=(10000, 10) M_\odot$ \\ $f_{\rm cut}=0.44$ Hz }}
 \\
\hline 
  & $\phi_{\rm circ} $ & $\phi (e^2)$ & $\phi (e^4)$ & $\phi (e^6)$ & $\phi (e^8)$ & $\phi (e^{10})$ \\
\hline
3.5PN  & 
\textbf{-182} & -6.75 & -6.56 & -6.77 & -6.77 & -6.77 \\ 
\hline
4PN  &  
\textbf{-305} & 4.95 & 6.02 & 6.05 & 6.06 & 6.06 \\ 
\hline
4.5PN & 
\textbf{232} & -1.50 & -1.73 & -1.78 & -1.78 & -1.78 \\ 
\hline
5PN &  
-256 & 0.611 & 0.777 & 0.795 & 0.798 & 0.799 \\ 
\hline
\hline
\end{tabular}
\end{center}
\end{table}

\begin{table*}[t]
\begin{center}
\captionsetup{justification=raggedright, singlelinecheck=false}
\caption{
Contributions from circular terms in the $\eta\rightarrow0$ limit of the phasing model for three representative IMRIs as a function of the PN order are shown. Choices for $e_0$ as well as that for the lower and upper frequency cut-off are same as in Table~\ref{table:GWcycles}. Other details are same as given in Table \ref{table:GWcycles}. We note that for $(m_1,m_2)=(10000,30)M_\odot$ the phase at 12PN is $\mathcal{O}(1)$. Thus, for the total mass $\sim 10^4M_\odot$, we would need higher PN order expressions than 12PN if we consider smaller mass ratios than $3\times 10^{-3}$ in DECIGO frequency band since the number of GW cycles would increase.}
\label{table:12PNorderphase} 
\begin{tabular}{|c|c|c|c|c|}
\hline
\hline
 & \multicolumn{3}{|c|}{$ N_{\rm gw}^{\rm BHP}$}  \\
 \hline
 PN order & \multicolumn{1}{|c|}{\shortstack{ $(m_1, m_2) = (100, 10) M_\odot$, \\ $f_{\rm cut}=40 $ Hz }} & \multicolumn{1}{|c|}{\shortstack{ $(m_1, m_2) = (1000, 10) M_\odot$ \\ $f_{\rm cut}=4.3 $ Hz }} & \multicolumn{1}{|c|}{\shortstack{ $(m_1, m_2) = (10000, 30) M_\odot$ \\ $f_{\rm cut}=0.44 $ Hz }}  
 \\
\hline
 0PN  &  1.07 $\times 10^7$ & 2.24 $\times 10^6$ & 1.60 $\times 10^5$ 
 \\ 
\hline
1PN  &  2.61 $\times 10^4$ & 2.39 $\times 10^4$ & 7.74 $\times 10^3$ 
\\ 
\hline
1.5PN  & -5.69 $\times 10^3$ & -1.08 $\times 10^4$  & -7.19 $\times 10^3$   
\\ 
\hline
 2PN  &  66.0 & 246  & 311  
 \\ 
\hline
2.5PN & -60.1 & -371  & -759    
\\ 
\hline
3PN   & -1.34 & -6.88 & -3.48    
\\ 
\hline
3.5PN  &  -2.38 & -19.8  & -60.9   
\\ 
\hline
4PN & -3.80 & -32  & -103 
\\ 
\hline
4.5PN   & 2.83 & 23.9  & 77.8 
\\ 
\hline
5PN  & -3.09  & -26.1 & -85.6 
\\ 
\hline
5.5PN & 2.89  & 24.4  & 80.1  
\\ 
\hline
6PN   & -2.43 & -20.5 & -67.3   
\\ 
\hline
 6.5PN  &  2.0 & 16.8  & 55.4 
 \\ 
\hline
 7PN  & -1.57 & -13.2 & -43.4  
 \\ 
\hline
 7.5PN  & 1.22 & 10.3 & 33.7 
 \\ 
\hline
 8PN  &  -0.893 & -7.53 & -24.7 
 \\ 
\hline
8.5PN   &  0.659 & 5.56 & 18.3   
\\ 
\hline
9PN   &  -0.465 & -3.92 & -12.9 
\\ 
\hline
9.5PN   & 0.328 & 2.76 & 9.08 
\\ 
\hline
10PN  & -0.223 & -1.88 & -6.17   
\\ 
\hline
 10.5PN  & 0.151 & 1.27  & 4.19 
 \\ 
\hline
 11PN  & -0.098 & -0.826  & -2.71  
 \\ 
\hline
 11.5PN  & 0.064 & 0.536 & 1.76   
 \\ 
\hline
 12PN  &  -0.039 & -0.327 & -1.08  
 \\ 
\hline
Total & 1.07 $\times 10^7$ & 2.25 $\times 10^6$ & 1.60 $\times 10^5$    
\\ 
\hline 
\hline
\end{tabular}
\end{center}
\end{table*}

\begin{itemize}
    \item In Table~\ref{table:GWcycles}, the low mass case, $(100, 10)M_{\odot}$ (major column 2) and the intermediate mass case, $(1000, 100)M_{\odot}$ (major column 3), attempt to highlight the effect of the additional mass ratio information available in PN part of the model by subtracting the number of cycles estimates keeping both results accurate through the 4.5PN(3PN) order for the circular (eccentric) part. For the circular part, 
    contributions from the $\eta$ dependent terms absent in BHP phase (included as bold text within the parenthesis for comparison) amount to nearly 10\% of the  contribution to the phase at 1PN order. Similar contributions at higher orders, though relatively smaller, still contribute significantly -- $\mathcal{O}(1-100)$ GW cycles. The dominance of the circular 1PN term clearly reflects in the total contribution from these terms shown in the bottom row. 
    \item In Table~\ref{table:GWcycles2}, the high mass case, $(10000, 10)M_{\odot}$, attempts to highlight the effect of the higher order contributions available only within the BHP approach beyond certain orders -- 4.5PN (3PN) for circular (eccentric) case.\footnote{We highlight in bold the contributions which can also be computed using PN results. These have been included to demonstrate that the number of cycles estimates for the circular case do not converge up to the 5PN order as discussed below.}As can be seen in the table, these contributions amount to over 250 cycles for the circular case through contributions at 5PN order. Contributions from eccentric BHP terms at orders beyond 3PN are of the order $\mathcal{O}(1)$ cycle and thus are in principle detectable and can induce systematic errors. We note that $N_{\rm gw}$ converges to a value smaller than 1 cycle at 5PN, indicating that for systems as heavy as the one under consideration, our 5PN eccentric model should be appropriate. On the other hand, contributions from circular terms up to 5PN neither have converged nor are anywhere close to contributing less than a cycle. We shall come back to this observation later in this section.  
    \item Each major column also compares (across rows) these estimates as a function of the eccentricity order. We find for each systems that the phasing results with leading eccentricity corrections significantly underestimate the number of GW cycles -- almost by a factor of 10 at 1PN independent of the mass of the binary. Naturally, number of cycles estimates from eccentric corrections reduce as move from low to intermediate to high mass system just like those from circular terms (see Table~\ref{table:GWcycles}-\ref{table:GWcycles2}).    
\end{itemize}

To address the concern raised above (in Table~\ref{table:GWcycles2}) that contributions from circular BHP terms do not converge through 5PN, we compute the number of cycles from such terms through 12PN order for three representative binaries. The estimates are presented in Table~\ref{table:12PNorderphase}. The highest mass system $(10000, 30)M_\odot$ is chosen in such a way that the contribution from the 12PN term is close to 1 cycle. We note that for each configuration the absolute $N_{\rm gw}$ estimates converge beyond 5PN order. Based on a $\mathcal{O}(1)$ cycle in the band criterion, these results indicate that our model (with circular contributions to the phase from BHP approach through 12PN) should be suitable for analyzing systems with mass ratios in the range, $0.1$--$3\times10^{-3}$ and total mass in the range $\sim100\,$--$10^{4}$ --- parameter space spanned by IMRIs involving at least an intermediate mass BH paired with an another BH of stellar or intermediate mass and observable in the deci Hz band.   

%%%%%%%%%%%%%%%%%%%%%%%%%%%%%%%%%%%%%%%%%
%%%%%%%%%%%%%%%%%%%%%%%%%%%%%%%%%%%%%%%%%

\section{Summary and Discussion}
\label{sec:concl}

Our focus through this exercise has been two-fold. One, to facilitate `direct' comparisons of results from the PN and BHP approaches by obtaining relations connecting the parametrizations adopted by the two approaches. The second goal is to provide a fully analytical, ready-to-use model suitable for exploring the eccentric, IMRI space. IMRIs with at least one IMBH as components is expected to merge in the dHz band, offering a new window to explore the compact binary dynamics.

Relations connecting the parametrizations of the two approaches were obtained in Sec.~\ref{subsec:BHPtoPNconv}; see Eqs.~\eqref{eq:vbvrelation}-\eqref{eq:eberelation}. These were subsequently used to re-express the inputs for the computation of the phase. Note that, once the inputs from the two approaches are written using the same parametrization (we choose to work with the one used within the PN approach) the two sets of results can simply be combined by simply switching on the contributions from the BHP approach at orders beyond which the information from the PN approach is not known yet. Naturally, the relations connecting the two approaches were only needed in the $\eta\rightarrow0$ limit. Also needed, were relations evolving orbital eccentricity as a function of time. For comparing the amplitude, additionally, we needed to establish connection between two set of phase angles appearing in the expression for spherical harmonic modes for eccentric systems. These were obtained by comparing orbit averaged radial and azimuthal frequencies from the two approaches. 

Our hybrid (PN-BHP) model is 3PN accurate in terms of results from PN theory and 5PN in results from BHP theory in both amplitude and phase. This is further complemented by computations of phase through 4.5PN in PN~\cite{Blanchet:2023bwj} and up to 12PN within the BHP approach~\cite{Fujita:2012cm} for the circular part. The model is presented in Sec.~\ref{sec:hybrid}. The time domain phasing is computed using the \TTtwo{} approximant, while the corresponding frequency domain phasing, \TFtwo{} obtained under the SPA approximation. Note that, while only ($\ell=2$, $|m|=2$) mode through the 5PN order is explicitly listed, expressions for all other modes (up to $\ell=|m|=12$) can be accessed through the \textbf{supplemental material}~\cite{supplementalfile}. Note also, only leading eccentricity results (for both amplitude and phase) are explicitly listed while the model is accurate through 10th power in eccentricity parameter and can be assessed via the \textbf{supplemental material}~\cite{supplementalfile}. 

Section~\ref{sec:GWcycles} attempts to highlight complementarity of the two approaches via numerical estimates for number of GW cycles in the dHz band. Table~\ref{table:GWcycles} and \ref{table:GWcycles2} presents these estimates for three representative IMRI configurations. We find that, completeness of mass ratio information in the PN approach and high PN results of the BHP help describing IMRI parameter space. Further, we also see that leading eccentricity corrections significantly underestimate the number of GW cycles, sometimes, by factor of almost 10. Additionally, the impact of complementing our results through 12PN order for the circular part employing the results from the BHP approach is demonstrated in Table~\ref{table:12PNorderphase}. These estimates suggest that our model should be adequate for analyzing systems with mass ratios in the range, $0.1$--$3\times10^{-3}$ and total mass in the range $\sim100\,$--$10^{4}M_\odot$ --- an IMRI parameter space.

Further, the zero frequency contributions  to $m=0$ modes are known through the 3PN order within the PN approach~\cite{Ebersold:2019kdc}, however, these are not computed within the BHP approach. We supplement this information from the PN approach when providing expressions for these modes as part of the \textbf{supplemental material}~\cite{supplementalfile}. Finally, horizon contributions at 4PN order and beyond is available from the BHP approach and have been incorporated in our model. Note for instance, that circular terms at 4PN order will only match with those of \cite{Blanchet:2023sbv,Blanchet:2023bwj} if the horizon contribution is subtracted. We demonstrate this in Appendix~\ref{sec:BHhorizonContributions}.

%%%%%%%%%%%%%%%%%%%%%%%%%%%%%%%%%%%%%%%%%
%%%%%%%%%%%%%%%%%%%%%%%%%%%%%%%%%%%%%%%%%

\acknowledgments
ML thanks members of the Gravitation and Cosmology group at the Department of Physics, Indian Institute of Technology (IIT) Madras for useful discussions.
CKM acknowledges the support of SERB’s Core Research Grant No.~CRG/2022/007959. RF's work was supported in part by JSPS/Ministry of Education, Culture, Sports, Science and Technology (MEXT) KAKENHI Grants No.~JP21H01082 and No.~JP23K20845. 

% \clearpage

%%%%%%%%%%%%%%%%%%%%%%%%%%%%%%%%%%%%%%%%%%%%%%%%%%%
%%%%%%%%%%%%%%%%%%%%%%%%%%%%%%%%%%%%%%%%%%%%%%%%%%%

\appendix
% \vspace{20pt}
\section{Enhancement functions}
\label{App:A}
As noted in Sec.~\ref{subsec:PNfluxandphase}, the RHS of Eq.~\eqref{eq:dv} depends explicitly on $e_t$ which must be substituted for via its expression in terms of $v$ that can be obtained by solving the evolution equation for $e_t$ and $v$; see for instance Sec. IV C of \cite{Moore:2016qxz} for a discussion. Equation (4.17) of \cite{Moore:2016qxz} provides 3PN expression for $e_t$ at the leading order in eccentricity parameter ($e_0$). We extend this to $\mathcal{O}(e_0^{10})$ and to 5PN where contributions beyond 3PN are from BHP approach; see Eq.~\eqref{eq:et2e0lead}. Evolution equation for $e_t$ is known through 3PN order within the PN approach and through 5PN in BHP approach. Note, however, not all contributions to these have closed-form expressions, although, an eccentricity expanded form can be obtained~\cite{Arun:2007rg,Arun:2007sg,Arun:2009mc}. (Typically, the equation can be split into an instantaneous part and a hereditary part that is expressed in terms of enhancement functions and not all of them take closed form expressions.) Closed-form expressions for enhancement functions associated with evolution equations for $e\equiv e_t$ to 3PN order, in an eccentricity expanded form to $\mathcal{O}(e_t^{6})$ are available in Ref.~\cite{Ebersold:2019kdc}; see Eqs.~B7(a)-(j) there. We update them to $\mathcal{O}(e_t^{10})$ following a prescription outlined in \textbf{Appendix C} of Ref.~\cite{Klein:2018ybm}; see Eqs.~C3(a)-(h) there.\footnote{Note however that, Ref.~\cite{Klein:2018ybm} recomputes the enhancement functions of~\cite{Arun:2007rg, Arun:2007sg, Arun:2009mc} following the method presented there to improve their convergence by means of absorbing the factor $\sqrt{1-e_t^2}$ into a new definition of frequency, $y$, related to the velocity parameter $v$ used in our work and $e_t$ as
\begin{align*}
    y=\frac{v}{\sqrt{1-e_t^2}}
\end{align*}}
All the relevant enhancement functions for the evolution equation of $e_t$ expanded through $\mathcal{O}(e_t^{10})$ (following the notation of Ref.~\cite{Ebersold:2019kdc}) read
\begin{widetext}
\begin{subequations}
\label{eq:App:Enhancement_fun}
    \begin{equation}
        \phi(e_t)=1+\frac{2335}{192}e_{t}^{2}+\frac{42955}{768}e_{t}^{4}+\frac{6204647}{36864}e_{t}^{6}+\frac{352891481}{884736}e_{t}^{8}+\frac{286907786543}{353894400}e_{t}^{10} + \mathcal{O}(e_t^{10})\,,
     \end{equation}

     \begin{equation}
        \tilde{\phi}(e_t)=1+\frac{209}{32}e_{t}^{2}+\frac{2415}{128}e_{t}^{4}+\frac{730751}{18432}e_{t}^{6}+\frac{10355719}{147456}e_{t}^{8}+\frac{6594861233}{58982400}e_{t}^{10} + \mathcal{O}(e_t^{10})\,,
     \end{equation}

     \begin{equation}
        \psi(e_t)=1-\frac{22988}{8191}e_{t}^{2}-\frac{36508643}{524224}e_{t}^{4}-\frac{1741390565}{4718016}e_{t}^{6}-\frac{749658956273}{603906048}e_{t}^{8}-\frac{8194146444139}{2516275200}e_{t}^{10}+ \mathcal{O}(e_t^{10}) \,,
     \end{equation}

     \begin{equation}
        \tilde{\psi}(e_t)=1-\frac{17416}{8191}e_{t}^{2}-\frac{14199197}{524224}e_{t}^{4}-\frac{467169215}{4718016}e_{t}^{6}-\frac{150176020037}{603906048}e_{t}^{8}-\frac{3861553027147}{7548825600}e_{t}^{10}+ \mathcal{O}(e_t^{10}) \,,
     \end{equation}

     \begin{align}\nonumber
        \kappa(e_t) &= 1+\bigg(\frac{62}{3}-\frac{4613840}{350283}\ln{2}+\frac{24570945}{1868176}\ln{3}    \bigg)e_{t}^{2}+\bigg(\frac{9177}{64}+\frac{271636085}{1401132}\ln{2}-\frac{466847955}{7472704}\ln{3}\bigg)e_{t}^{4}+\bigg(\frac{76615}{128} \nonumber \\
        & -\frac{4553279605}{2802264}\ln{2}+\frac{14144674005}{119563264}\ln{3}+\frac{914306640625}{1076069376}\ln{5}     \bigg)e_{t}^{6}+\bigg(\frac{1903055}{1024}+\frac{135819536755}{22418112}\ln{2} \nonumber \\
        &+\frac{3964202553465}{956506112}\ln{3}-\frac{39315185546875}{8608555008}\ln{5}\bigg)e_{t}^{8}+\bigg(\frac{9732723}{2048}-\frac{61237079991493}{2017630080}\ln{2}-\frac{3772118655280899}{153040977920}\ln{3} \nonumber \\
        &+\frac{8770943603515625}{826421280768}\ln{5}+\frac{24891464996631149}{1377368801280}\ln{7}\bigg)e_{t}^{10} + \mathcal{O}(e_t^{10}) \,,
    \end{align}

    \begin{align}\nonumber
    \tilde{\kappa}(e_t) &= 1+\bigg(\frac{389}{32}-\frac{2056005}{233522}\ln{2}+\frac{8190315}{934088}\ln{3}    \bigg)e_{t}^{2}+\bigg(\frac{3577}{64}+\frac{50149295}{467044}\ln{2}-\frac{155615985}{3736352}\ln{3}\bigg)e_{t}^{4}+\bigg(\frac{43049}{256} \nonumber \\
    & -\frac{12561332945}{16813584}\ln{2}+\frac{4709431125}{59781632}\ln{3}+\frac{182861328125}{538034688}\ln{5}     \bigg)e_{t}^{6}+\bigg(\frac{102005}{256}+\frac{43574367025}{16813584}\ln{2} \nonumber \\
     &+\frac{642112505685}{478253056}\ln{3}-\frac{7863037109375}{4304277504}\ln{5}\bigg)e_{t}^{8}+\bigg(\frac{207311}{256}-\frac{95758853813}{9340880}\ln{2}-\frac{626952624117057}{76520488960}\ln{3} \nonumber \\
    &+\frac{584607666015625}{137736880128}\ln{5}+\frac{3555923570947307}{688684400640}\ln{7}\bigg)e_{t}^{10} + \mathcal{O}(e_t^{10}) \,,
    \end{align}

    \begin{equation}
    \zeta(e_t)=1+\frac{1011565}{48972}e_{t}^{2}+\frac{106573021}{783552}e_{t}^{4}+\frac{456977827}{854784}e_{t}^{6}+\frac{128491074157}{82059264}e_{t}^{8}+\frac{342306246988373}{90265190400}e_{t}^{10} + \mathcal{O}(e_t^{10}) \,,
     \end{equation}

     \begin{equation}
        \tilde{\zeta}(e_t)=1+\frac{102371}{8162}e_{t}^{2}+\frac{14250725}{261184}e_{t}^{4}+\frac{722230667}{4701312}e_{t}^{6}+\frac{102744533069}{300883968}e_{t}^{8}+\frac{9843430194463}{15044198400}e_{t}^{10} + \mathcal{O}(e_t^{10}) \,,
     \end{equation}

     \begin{equation}
        F(e_t)=1+\frac{62}{3}e_{t}^{2}+\frac{9177}{64}e_{t}^{4}+\frac{76615}{128}e_{t}^{6}+\frac{1903055}{1024}e_{t}^{8}+\frac{9732723}{2048}e_{t}^{10}+ \mathcal{O}(e_t^{10}) \,,
     \end{equation}

     \begin{equation}
        \tilde{F}(e_t)=1+\frac{389}{32}e_{t}^{2}+\frac{3577}{64}e_{t}^{4}+\frac{43049}{256}e_{t}^{6}+\frac{102005}{256}e_{t}^{8}+\frac{207311}{256}e_{t}^{10}+ \mathcal{O}(e_t^{10}) \,.
     \end{equation}    
\end{subequations}
\end{widetext}

\section{Differential equation for the evolution of PN variables}
\label{App:PN_Evolution_eqn}

Enhancement functions computed in Appendix~\ref{App:A} can further be used in explicitly write evolution equations for orbital eccentricity ($e_t$) and that of frequency dependent PN parameter ($v$) as follows. 

\begin{widetext}
\begin{subequations}
\begin{align}
\label{eq:App:Hybrid_dedt_diffeqn}
    \bigg(\frac{de_t}{dt}\bigg)^{\rm Hybrid} =\bigg[\bigg(\frac{de_t}{dt}\bigg)_{e_t}+\sum^{4}_{j=1} \bigg(\frac{de_t}{dt}\bigg)_{e_t^{2j+1}}\bigg] + \mathcal{O}(e_t^{9}) \,,
\end{align}

\begin{align} \nonumber
\label{eq:App:Hybrid_leading_dedt_diffeqn}
    \bigg(\frac{de_t}{dt}\bigg)_{e_t} &=  -\frac{304\,e_t \eta v^8}{15 m}\bigg[1-\bigg(\frac{2817}{2128}+\frac{1021}{228}\eta \bigg) v^2 + \frac{985}{152} \pi v^3 + \bigg(-\frac{108197}{38304}+\frac{56407}{4256} \eta+\frac{141}{19}\eta^2\bigg) v^4 - \bigg(\frac{55691}{4256} \\ \nonumber
    &+\frac{19067}{399}\eta \bigg)\pi v^5 + \bigg(\frac{246060953209}{884822400}-\frac{82283}{1995}\gamma_E +\frac{769}{57}\pi^2+\bigg(-\frac{613139897}{2298240}+\frac{22345}{3648}\pi^2 \bigg) \eta -\frac{1046329}{51072}\eta^2 \\ \nonumber
    & -\frac{305005}{49248} \eta^3-\frac{11021 }{285}\ln 2-\frac{234009}{5320}\ln
    3-\frac{82283}{1995} \ln v \bigg)v^6 -\frac{195209}{21888}\pi v^7+ \bigg(-\frac{5266265642641}{32207535360}\\ \nonumber
    & +\frac{730568}{4655} \gamma_E-\frac{4329}{133} \pi^2-\frac{10096}{133}\ln 2 +\frac{43300413}{148960}\ln 3 +\frac{730568}{4655} \ln v \bigg) v^8 + \bigg(\frac{4858043556367}{2831431680}-\frac{202123}{798}\gamma_E \\ \nonumber
    & -\frac{969527}{3990}\ln 2 -\frac{702027}{2660}\ln 3 -\frac{202123}{798} \ln v \bigg)\pi v^9 +\bigg(\frac{51696376215952981}{42513946675200}-\frac{2094212137 }{11060280}\gamma_E-\frac{175825}{9576}\pi^2  \\ 
    & +\frac{18659835527}{11060280}\ln 2-\frac{328813263}{344960}\ln 3 -\frac{76708984375}{318536064}\ln 5 -\frac{2094212137}{11060280}\ln v \bigg)v^{10} + \mathcal{O}(v^{10})\bigg] + \mathcal{O}(e_t)\,.
\end{align}    
\end{subequations}

\begin{subequations}
\begin{align}
\label{eq:App:Hybrid_dvdt_diffeqn}
    \bigg(\frac{dv}{dt}\bigg)^{\rm Hybrid} =\bigg[\bigg(\frac{dv}{dt}\bigg)_{\mathrm{circ}}+\sum^{5}_{j=1} \bigg(\frac{dv}{dt}\bigg)_{e_t^{2j}}\bigg] + \mathcal{O}(e_t^{10}) \,,
\end{align}

\begin{align} \nonumber
\label{eq:App:Hybrid_leading_dvdt_diffeqn}
    \bigg(\frac{dv}{dt}\bigg)_{\rm circ} &= \frac{32  v^9 \eta}{5 m}\bigg[1 
    - \bigg(\frac{743}{336} + \frac{11}{4}\eta \bigg) v^2
    + 4 \pi v^3
    + \bigg(\frac{34103}{18144} + \frac{13661}{2016}\eta + \frac{59}{18}\eta^2 \bigg) v^4
    - \bigg(\frac{4159}{672} + \frac{189}{8}\eta\bigg) \pi v^5 \\ \nonumber
    & + \bigg(\frac{16447322263}{139708800} - \frac{1712}{105} \gamma_E + \frac{16}{3} \pi^2 
    + \bigg(-\frac{56198689}{217728} + \frac{451}{48} \pi^2\bigg)\eta 
    + \frac{541}{896}\eta^2 - \frac{5605}{2592}\eta^3 - \frac{3424}{105}\ln 2 \\ \nonumber
    & - \frac{1712}{105} \ln v \bigg) v^6 -\frac{4415}{4032} \pi v^7 + \bigg(\frac{3984698178313}{25427001600} + \frac{124741}{4410}\gamma_E - \frac{361}{126} \pi^2 
     + \frac{127751}{1470} \ln 2 - \frac{47385}{1568} \ln 3 \\ \nonumber
    & + \frac{124741}{4410} \ln v \bigg) v^8 + \bigg(\frac{343801320119}{745113600} - \frac{6848}{105} \gamma_E - \frac{13696}{105} \ln 2  - \frac{6848}{105} \ln v\bigg)\pi v^9  \\ \nonumber
    &  + \bigg(\frac{29818518973687069}{36248733480960} - \frac{11821184}{1964655}\gamma_E - \frac{21512}{1701} \pi^2 - \frac{107638990 }{392931}\ln 2 + \frac{616005}{3136} \ln 3 \\ 
    & - \frac{11821184}{1964655} \ln v \bigg) v^{10} + \mathcal{O}(v^{10}) \bigg] + \mathcal{O}(e_t^{}) \,.
\end{align}
\end{subequations}

These can then be combined to obtain the equation which can be solved to obtain Eq.~\eqref{eq:str_et2e0}. 

\begin{subequations}
\begin{align}
\label{eq:App:Hybrid_dedv_diffeqn}
    \bigg(\frac{de_t}{dv}\bigg)^{\rm Hybrid} = \bigg[\bigg(\frac{de_t}{dv}\bigg)_{e_t}+\sum^{4}_{j=1} \bigg(\frac{de_t}{dv}\bigg)_{e_t^{2j+1}}\bigg] + \mathcal{O}(e_t^{9})\,,
\end{align}

\begin{align} \nonumber
\label{eq:App:Hybrid_leading_dedv_diffeqn}
    \bigg(\frac{de_t}{dv}\bigg)_{e_t} &= -\frac{19 e_t}{6v}\bigg[1 + \bigg(\frac{2833}{3192} - \frac{197 \eta}{114}\bigg) v^2 + \frac{377 \pi}{152} v^3 + \bigg(-\frac{1392851}{508032} + \frac{32537}{6384} \eta - \frac{833}{1368} \eta^2 \bigg) v^4 + \bigg(-\frac{253409}{51072} \\ \nonumber
    & - \frac{133157}{12768} \eta\bigg)\pi v^5+ \bigg(\frac{27226918334431}{178380195840} - \frac{3317}{133} \gamma_E - \frac{67}{38} \pi^2 
    + \bigg(-\frac{26105879}{3386880} - \frac{3977}{1216} \pi^2\bigg)\eta 
    + \frac{58057}{153216} \eta^2  \\ \nonumber
    & - \frac{25}{608} \eta^3 - \frac{12091}{1995} \ln 2 - \frac{234009}{5320} \ln 3 - \frac{3317}{133} \ln v \bigg) v^6
    - \frac{360256021}{51480576} \pi v^7  
    + \bigg(-\frac{2876250907217867303}{35062411294310400} 
    \\ \nonumber
    &  + \frac{58973387}{670320} \gamma_E - \frac{318553}{102144} \pi^2 
    - \frac{134315}{912} \ln 2 + \frac{133251723}{595840} \ln 3 
    + \frac{58973387}{670320} \ln v\bigg) v^8 + \bigg(\frac{20909324518837}{63424069632} \\ \nonumber
    & - \frac{1819}{38} \gamma_E - \frac{352}{57} \pi^2  - \frac{29639}{3990} \ln 2 - \frac{234009}{2660}\ln 3 - \frac{1819}{38}\ln v\bigg)\pi v^9
    + \bigg(\frac{522055034134608725351}{4799654523843379200} \\ \nonumber
    & - \frac{783468941251}{66892573440} \gamma_E + \frac{57859117}{7354368} \pi^2 
    + \frac{99016182356381}{66892573440} \ln 2 - \frac{400507662699 }{734074880}\ln 3 - \frac{76708984375}{318536064} \ln 5 
    \\ 
    & - \frac{783468941251}{66892573440} \ln v\bigg) v^{10} + \mathcal{O}(v^{10}) \bigg] + \mathcal{O}(e_t^{})\,.
\end{align}
\end{subequations}
\end{widetext}

\textcolor{black}{The full expression for the 
$dv/dt$, $de_t/dt$ and $de_t/dv$ accurate up to tenth order in eccentricity is included in the \textbf{supplemental material}~\cite{supplementalfile}}

%%%%%%%%%%%%%%%%%
%%%%%%%%%%%%%%%%%

\section{Equivalence of the radial frequency variables in PN and BHP theory}
\label{App:B}

We show the equivalence of the orbit averaged radial frequency in PN ($\omega_r$) and BHP ($\Omega_r$) approach here. This can be verified by using Eqs.~\eqref{eq:vbvrelation} and \eqref{eq:eberelation} obtained earlier in Sec.~\ref{subsec:BHPtoPNconv} in the 3PN limit. Through 3PN order, $\Omega_r$ reads

\begin{widetext}    
\begin{align}
\label{eq:App:BHP_Radialfreq}
    \Omega_{r} &= v_{b}^{3}\bigg[\bigg(1-\frac{3}{2}e_{ b}^{2}+\frac{3}{8}e_{b}^{4}+\frac{1}{16}e_{b}^{6}+\frac{3}{128}e_{b}^{8}+\frac{3}{256}e_{b}^{10}\bigg)+\bigg(-3+\frac{15}{2}e_{ b}^{2}-\frac{45}{8}e_{b}^{4}+\frac{15}{16}e_{b}^{6}+\frac{15}{128}e_{b}^{8}+\frac{9}{256}e_{b}^{10}\bigg)v_{b}^{2} \nonumber \\
    & +\bigg(-\frac{9}{2}+6\,e_{b}^{2}+\frac{45}{8}e_{b}^{4}-\frac{165}{16}e_{b}^{6}+\frac{345}{128}e_{b}^{8}+\frac{81}{256}e_{b}^{10}\bigg)v_{b}^{4}+\bigg(-\frac{27}{2}+9\,e_{b}^{2}+\frac{249}{8}e_{b}^{4}-\frac{375}{16}e_{b}^{6}-\frac{1575}{128}e_{b}^{8} \nonumber \\
    & +\frac{1983}{256}e_{b}^{10}\bigg)v_{b}^{6} +\mathcal{O}(v_b^{6})+\mathcal{O}(e_b^{10}) \bigg] \,.
\end{align}
\end{widetext}

Substituting for $v_b$ in terms of $(v, e_b)$ from Eq.~\eqref{eq:vbvrelation} and subsequently $e_b$ in terms of $(v, e_t)$ from Eq.~\eqref{eq:eberelation} we get

\begin{widetext}
\begin{align}
\label{eq:App:BHPradialfreq_InPN}
    \Omega_{r} &= v^{3}\bigg[1-3(1+e_t^{2}+e_t^{4}+e_t^{6}+e_t^{8}+e_t^{10})v^{2}-\bigg(\frac{9}{2}+\frac{87}{4}e_t^{2}+39 e_t^{6}+\frac{225}{4}e_t^{6}+\frac{147}{2}e_t^{8}+\frac{363}{4}e_t^{10}\bigg)v^{4} \nonumber \\
    & -\bigg(\frac{27}{2}+\frac{519}{4}e_t^{2}+\frac{2811}{8}e_t^{6}+\frac{10779}{16}e_t^{6}+\frac{140061}{128}e_t^{8}+\frac{412521}{256}e_t^{10}\bigg)v^{6} +\mathcal{O}(v^{6})+\mathcal{O}(e_t^{10}) \bigg]\,.
\end{align}
\end{widetext}

As can be verified, up to 3PN order RHS of the above equation is same as the expression for the radial frequency variable in PN theory given by Eq~(3.10) of Ref.~\cite{Moore:2016qxz}. And thus the orbit averaged radial frequencies of the two approaches can be treated equivalent.  

%%%%%%%%%%%%%%%%%
%%%%%%%%%%%%%%%%%

\section{BH horizon contribution}
\label{sec:BHhorizonContributions}
We highlighted in Sec.~\ref{sec:hybrid} that our results include the BH horizon contributions available at orders
4PN and beyond in the test particle limit and special care must be taken while comparing the circular limit of our results with that of Ref.~\cite{Blanchet:2023sbv, Blanchet:2023bwj} which does not compute these contributions. Here, we recover the results of Ref.~\cite{Blanchet:2023sbv, Blanchet:2023bwj} staring from results reported in this work. Note that Ref.~\cite{PhysRevD.107.084006} explicitly lists this contribution for the energy flux ($\mathcal{F}^{\rm H}$) in the test particle limit for binaries in circular orbit. For nonspinning binaries, the 4PN contribution to the energy flux reads,

\begin{widetext}
 \begin{align}
 \label{eq:App:Horizonflux}
     \mathcal{F}^{\rm H}_{\rm circ} &= \frac{32}{5}\eta^2 v^{10}\bigg[v^8 + \mathcal{O}(v^{9})\bigg].
\end{align} 
\end{widetext}

Subtracting this from the circular part of Eq.~\eqref{eq:Hybridflux_circ_ecc} we recover the flux of \cite{Blanchet:2023sbv} --- $\mathcal{F}^{\infty}_{\rm circ}$. Naturally, removing the horizon contribution at 4PN precisely alters only one term displayed below within the square brackets. 

\begin{widetext}

\begin{align}
\label{eq:App:Flux_infinity}
    \mathcal{F}^{\infty}_{\rm circ} &= \frac{32}{5}\eta^2 v^{10}\bigg\{1+ \cdots + \bigg[-\frac{323105549467}{3178375200} + \cdots \bigg]v^8 + \mathcal{O}(v^{9}) \bigg\}.  
\end{align}
\end{widetext}

Further, employing this flux together with the expression for energy we reproduce the expression for the \TTtwo{} and \TFtwo{} phase consistent with those of \cite{Blanchet:2023bwj} as shown below.

\begin{widetext}
\begin{align}
\label{eq:App:phase_infinity}
     \langle \phi\rangle^{\infty}_{\mathrm{circ}} - \phi_c &= -\frac{1}{32v^5\eta} \bigg\{1+\cdots + \bigg[\frac{2550713843998885153}{2214468081745920} + \cdots  
     \bigg] v^8
     + \mathcal{O}(v^{9}) \bigg\} \,,
\end{align}

\begin{align}
\label{eq:App:SPAphase_infinity}
     \Psi^{\infty}_{\mathrm{SPA, circ}} &= \psi_0 + 2\pi f t_c+\frac{3}{128 }\frac{1}{v^5\eta} \bigg\{1+\cdots +\bigg\{-\frac{2550713843998885153}{276808510218240}+\cdots \bigg\}v^8 \ln v + 
       \mathcal{O}(v^{9}) \bigg\} \,.
\end{align}
\end{widetext}

Above expressions matches \TTtwo{} and \TFtwo{} phase computed in Ref.~\cite{Blanchet:2023bwj} in terms of PN parameter $x\equiv v^2$.

%%%%%%%%%%%%%%%%%%%%
%%%%%%%%%%%%%%%%%%%%

\section{Evolution of $W(l)$ function and connecting the redefined azimuthal and radial phase angles in PN theory}
\label{App:phitopsi_ltoxi_gen_eqn}

Following \cite{Munna:2020iju}, 5PN accurate expression for $W(l)$ in $\eta\rightarrow0$ limit can be given in the following form

\begin{widetext}
\begin{align} \nonumber \label{eq:WinV}
    W(l) &= (1+k)(\nu-l)+\bigg[e_t^2 \bigg(\frac{1}{8}v^4+\frac{21}{8} v^6+\frac{559}{16} v^8+\frac{6037}{16} v^{10} \bigg)+e_t^4\bigg(\frac{1}{4}v^4+\frac{65}{8} v^6+\frac{4907}{32}v^8+\frac{70707}{32} v^{10} \bigg)+e_t^6\bigg(\frac{3}{8} v^4 \\ \nonumber
    & +\frac{33}{2} v^6+\frac{25803}{64} v^8+\frac{232617}{32} v^{10}\bigg)+e_t^8\bigg(\frac{1}{2}v^4+\frac{111}{4}v^6+\frac{106509}{128} v^8+\frac{3457979}{192} v^{10}\bigg)+e_t^{10} \bigg(\frac{5}{8} v^4+\frac{335}{8} v^6  \\ \nonumber
    & +\frac{1523355}{1024} v^8+\frac{3605855}{96} v^{10} \bigg) \bigg]\sin (2\nu) + \bigg[e_t^4\bigg(\frac{3}{256} v^8+\frac{117}{256}v^{10} \bigg)+e_t^6\bigg(\frac{3}{64} v^8+\frac{597}{256} v^{10}\bigg)+e_t^8\bigg(\frac{15}{128} v^8+\frac{1815}{256} v^{10}\bigg) \\
    & +e_t^{10}\bigg(\frac{15}{64} v^8+\frac{4275}{256}v^{10} \bigg)   \bigg]\sin(4\nu)  
\end{align}
\end{widetext}
where $\nu$ is the true anomaly and through 5PN it reads

\begin{widetext}
\begin{subequations}
\begin{align}
\label{eq:Vinl_gen}
    \nu &= l+\nu_{e_{t}} + \sum_{j=2}^{10} \nu_{e_{t}^{j}}+\mathcal{O}(e_t^{10}) 
\end{align}
\begin{align}
\label{eq:Vinl}
    \nu_{e_t} &= 2e_t\bigg[1+2 v^2+\frac{13}{2} v^4+32 v^6+\frac{1447}{8} v^8+\frac{4169}{4}v^{10} +\mathcal{O}(v^{10}) \bigg]\sin(l) \,.
\end{align}
\end{subequations}
\end{widetext}

Ref.~\cite{Boetzel:2019nfw} computes a 3PN and leading order in eccentricity expression to connect the phase angles $(\phi, l)$ and $(\psi,\xi)$ and the same prescription can be used to extend it up to 5PN order in the test mass limit. We have 
\begin{align}
\label{eq:phitopsi_geneqn}
    \phi &= \psi - \sum_{s=1}^{s=\infty} \frac{1}{s!}\bigg[(\xi-l)^{s}(D_{l}^{s}\phi) + (\lambda_{\xi}-\lambda)^{s} (D_{\lambda}^{s}\phi) \bigg]\,,
\end{align}
where
\begin{subequations}
\label{eq:xitoell_lambdaxitolambda}
\begin{align}
\label{eq:xitoell_gen}
        \xi - l &=  -6M
        n \ln \bigg(\frac{v}{v_0^{\prime}}\bigg)\,,
\end{align}
and 
\begin{align}
\label{eq:lambdaxitolambda_gen}
    \lambda_{\xi} - \lambda &= - 6M
    (1+k)n \ln \bigg(\frac{v}{v_0^{\prime}}\bigg)\,.
\end{align}
\end{subequations}
Operators, $D_{l}^{s}$ and $D_{\lambda}^{s}$ are the $s^{\rm th}$ order derivative operator with respect to $l$ and $\lambda$ respectively. Angle $\lambda_{\xi}$ is the secular phase evaluated at the redefined phase angle $\xi$. Note that, $M=m(1-\eta v^2/2)$ is the ADM mass which is equal to the total mass of the binary $(m)$ in the test mass limit. Note also that summing Eq.~(\ref{eq:phitopsi_geneqn}) up to $s=3$ is adequate to obtain the conversion relation to 5PN accuracy. Finally, expressions for mean motion $n$ and $k$ with accuracies required for computing Eq.~\eqref{eq:phi2psi_l2xi} read
\begin{subequations}
\begin{align}
\label{eq:meanmotion_gen}
    n &= n_{0} + \sum_{j=1}^{5} n_{e_{t}^{2j}} +\mathcal{O}(e_t^{10})\,, 
\end{align}
\begin{align}
\label{eq:meanmotion}
    n_{0} &=  v^{3}\bigg[1-3v^2 -\frac{9}{2} v^4 -\frac{27}{2}v^6 -\frac{405}{8}v^8 - \frac{1701}{8}v^{10} \nonumber \\ 
    & + \mathcal{O}(v^{10}) \bigg]\,, 
\end{align}    
\end{subequations}
And,
\begin{subequations}
\begin{align}
\label{eq:precession_parameter_gen}
    k &= k_{0} + \sum_{j=1}^{5} k_{e_{t}^{2j}}+\mathcal{O}(e_t^{10})\,.
\end{align}
\begin{align} 
\label{eq:precession_parameter}
   k_{0} &= \bigg[3\bigg(v^2+\frac{9}{2} v^4+\frac{45}{2}v^6 + \frac{945}{8} v^8 +\frac{5103}{8}v^{10}  \bigg) \nonumber \\ 
   & + \mathcal{O}(v^{10}) \bigg]\,.
\end{align}
\end{subequations}

Note also, substituting, Eq.~\eqref{eq:Vinl_gen} for $\nu$ and Eq.~\eqref{eq:precession_parameter_gen} for $k$ in Eq.~\eqref{eq:WinV}, one can obtain the 5PN and leading order in eccentricity expression for $W(l)$ function in test mass limit given by Eq.~\eqref{eq:W(l)_function} in Sec.~\ref{sec:pn_bh_amp_relations}. The full expression for the 
$\nu$, $n$ and $k$ accurate up to tenth order in eccentricity is included in the \textbf{supplemental material}~\cite{supplementalfile}

\bibliographystyle{apsrev4-1}
\bibliography{master}

\end{document}